\documentclass[12pt,preprint]{aastex}

\shorttitle{BCMs and the formation of clusters}
\shortauthors{Coziol et al.}

\begin{document}

\title{The dynamical state of brightest cluster galaxies and the formation of clusters}

\author{R. Coziol}
\affil{Departamento de Astronom\'{\i}a, Universidad de Guanajuato\\
Apartado Postal 144, 36000 Guanajuato, Gto, M\'exico}
\email{rcoziol@astro.ugto.mx}
\author{H. Andernach}
\affil{Departamento de Astronom\'{\i}a, Universidad de Guanajuato\\
Apartado Postal 144, 36000 Guanajuato, Gto, M\'exico}
\email{heinz@astro.ugto.mx}
\author{C. A. Caretta}
\affil{Departamento de Astronom\'{\i}a, Universidad de Guanajuato\\
Apartado Postal 144, 36000 Guanajuato, Gto, M\'exico}
\email{caretta@astro.ugto.mx}
\author{K. A. Alamo-Mart\'{\i}nez}
\affil{Departamento de Astronom\'{\i}a, Universidad de Guanajuato\\
Apartado Postal 144, 36000 Guanajuato, Gto, M\'exico}
\email{k.alamo@astrosmo.unam.mx}
\author{E. Tago}
\affil{Tartu Observatory\\
61602 T\~oravere, Estonia}
\email{erik@aai.ee}

\begin{abstract}
A large sample of Abell clusters of galaxies, selected for the
likely presence of a dominant galaxy, is used to study the dynamical
properties of the brightest cluster members (BCMs). From visual
inspection of Digitized Sky Survey images combined with redshift
information we identify 1426 candidate BCMs located in 1221
different redshift components associated with 1169 different Abell
clusters. This is the largest sample published so far of such
galaxies. From our own morphological classification we find that
$\sim$ 92\% of the BCMs in our sample are early-type galaxies, and
48\% are of cD type.

We confirm what was previously observed based on much smaller
samples, namely that a large fraction of BCMs have significant
peculiar velocities. From a subsample of 452 clusters having at
least 10 measured radial velocities, we estimate a median BCM
peculiar velocity of 32\% of their host clusters' radial velocity
dispersion. This suggests that most BCMs are not at rest in the
potential well of their clusters. This phenomenon is common to
galaxy clusters in our sample, and not a special trait of clusters
hosting cD galaxies.

We show that the peculiar velocity of the BCM is independent of
cluster richness and only slightly dependent on the Bautz-Morgan
type. We also find a weak trend for the peculiar velocity to rise
with the cluster velocity dispersion. The strongest dependence is
with the morphological type of the BCM: cD galaxies tend to have
lower relative peculiar velocities than elliptical galaxies. This
result points to a connection between the formation of the BCMs and
that of their clusters.

Our data are qualitatively consistent with the merging-groups
scenario, where BCMs in clusters formed first in smaller subsystems
comparable to compact groups of galaxies. In this scenario, clusters would
have formed recently from the mergers of many such groups and would
still be in a dynamically unrelaxed state.

\end{abstract}

\keywords{galaxies: clusters: general -- galaxies: formation --
cosmology: large-scale structure of the Universe}

\section{Introduction}
Clusters of galaxies are dynamical systems formed by hundreds to
thousands of galaxies and an even larger mass of intergalactic hot
gas ($kT=0.5-15$\,keV; Hartley et al. 2008). Unfortunately,
understanding the formation of these structures turned out to be
much more complicated than originally expected. In part, this is
because we still do not know what is the nature of the main
constituent of these systems. The present paradigm for clusters of
galaxies states that both galaxies and gas are located in formidable
potential wells formed by non-baryonic dark matter halos.

Observational evidence in favor of this paradigm is based on various
mass estimates for clusters of galaxies (see Biviano et al. 2006 and
references therein). Considering first the velocity dispersions of
the galaxies in clusters, which range from a few hundred to over
1000~km\,s$^{-1}$, and assuming clusters to be dynamically relaxed,
the virial theorem suggests masses in the range of
$10^{13}-10^{15}$M$_\odot$. This exceeds a few hundred times the
mass deduced from the light of the galaxy members. We know now that
part of this missing mass is associated with the hot gas component.
In fact, X-ray observations suggest the bulk of the baryon mass in
clusters of galaxies is really in the form of gas, with an estimated
gas--to--stellar mass ratio of order $\sim 10:1$ (Mushotzky 2004).
If one assumes hydrostatic equilibrium, the corresponding dynamical
mass, once again, surpasses that of the directly observable mass,
i.e.\ gas and stars. According to this interpretation, no more than
10\% to 20\% of the mass in clusters is in the form of baryonic
matter.

According to the Cold Dark Matter (CDM) model, huge cluster halos
form hierarchically by the merging of smaller mass halos (see Loeb
2008 and references therein). Numerous simulations (White \& Rees
1978; Navarro, Frenk \& White 1995b; 1997) have shown that the final
dynamical state expected at $z = 0$ is one of a dynamically relaxed
system. But what about the galaxies that are observed within the
cluster? If these objects participate in the formation of the
cluster, then theoretical considerations and numerical simulations
suggest that through dissipation and cooling they would follow the
potential of the dark matter and settle into a dynamically relaxed
distribution (White \& Rees 1978; Navarro, Frenk \& White 1995a,
1995b, 1997; Benson et al. 2001; Springel et al. 2005; Loeb 2008).
According to this scenario, the BCMs would thus be expected to be at
rest at the bottom of the potential well of their clusters (Ostriker
\& Tremaine 1975; Hausman \& Ostriker 1978; Merritt 1984; Malumuth
1992).

One way to test the above prediction is to determine how far the
BCMs are located from the peak of the galaxy surface density in
their clusters. This was done by Beers \& Geller (1983). Using a
sample of 55 rich clusters of galaxies, these authors found cD
galaxies to lie at the bottom of local potential wells, rather than
global ones as would have been expected. Consistent with this
result, Malumuth et al. (1992), Zabludoff et al. (1993), Bird (1994)
and Oegerle \& Hill (2001), also found large fractions of D and cD
galaxies with significant peculiar velocities with respect to the
cluster mean. These results suggest BCMs are not at rest at the
bottom of the potential well of their clusters. However, the
question remains if these observations reflect a general pattern in
the formation of clusters or some special conditions connected only
to the formation of D and cD galaxies at the epoch of cluster
formation (Tonry 1985b).

Indeed, it is often assumed that BCMs are of D or cD type, i.e.\
huge elliptical galaxies with extended envelopes. The most probable
scenario for their formation is that they grow {\it in situ} through
the process of ``cannibalism'', which describes the accretion of
smaller mass companion galaxies (Hausman \& Ostriker 1978; Richstone
\& Malumuth 1983; Dubinski 1998). However, the accretion rates
possible in clusters today seem too low to produce the luminosity of
a D or cD galaxy in a reasonably short period of time (Hill et al.
1988; Lauer 1988; Merritt 1984; Tonry 1985b; Malumuth 1992; Nipoti
et al.\ 2003). The difficulty is due mostly to the velocities of
galaxies in clusters, which on average are too high for mergers to
be efficient (Mihos 2004).

The formation of BCMs in clusters is a more complicated matter. The
main question is how can we explain, at the same time, the formation
of a high number of galaxies (especially very massive giant Es, Ds
and cDs), and the predominance of a very hot gas component? Indeed,
the formation of a large number of galaxies in a system implies a
high level of efficiency for the process of transformation of gas
into stars (Gunn \& Gott 1972). This, in particular, requires very
short cooling times for the gas. Consequently, there should not be
any intergalactic hot gas left in clusters (Gunn \& Gott 1972; White
\& Rees 1978; Navarro, Frenk \& White 1995a; Suginohara \& Ostriker
1998). To solve this very serious difficulty, intricate and
contrived models were devised. Usually these include dynamical,
thermal and chemical inputs of star formation and AGNs on the
intracluster medium to eliminate the extra cooling, (e.g., Pearce et
al.\ 2000; Kay et al.\ 2003; Romeo et al.\ 2006; Domainko et al.\
2006). As ingenious as the models are, it seems difficult to
understand how these sporadic events could have left over such huge
amounts of gas after the formation of so many massive galaxies.

Considering the difficulties to explain how BCMs appear in clusters,
some authors suggested they really formed in smaller mass
subsystems, like compact groups of galaxies (Merritt 1985; Zabludoff
\& Mulchaey 1998). This hypothesis would easily solve the problem of
the low merger rates: compact groups have lower velocity
dispersions, favoring galaxy interactions and mergers (Merritt 1985;
Tonry 1985b; Mihos 2004; Coziol \& Plauchu-Frayn 2007), and
generally, very low quantities of intergalactic hot gas to inhibit
fast cooling (Ponman et al.\ 1996; Helsdon \& Ponman 2000; Mulchaey
et al.\ 2003; Jeltema et al.\ 2006).

Implicit in this scenario is that clusters must have formed from the
combination of many such groups. Now, assuming the BCMs retain part
of the infall velocity of their original groups, and assuming that
the clusters had not had enough time to relax dynamically since
their formation, the BCMs in such clusters would not be expected to
reside at the center of global potential wells, but only of local
ones, i.e.\ the potential well of their original groups. This would
explain their non-zero peculiar velocity (Malumuth 1992).

Consistent with the above scenario, the detection of substructures
in a significant number of clusters (Dressler \& Schectman 1988;
Schuecker et al.\ 2001; Flin \& Krywult 2006) suggests they are not
as relaxed as previously believed. Even the Coma cluster (A1656),
once considered the ``prototype of a rich, relaxed cluster'', is now
recognized as a relatively young, and still evolving dynamical
system (Ledlow et al.\ 2003; Adami et al.\ 2005). Reinforcing this
view, X-ray studies have recently revealed gas substructures,
turbulence, and even cases of mergers between clusters (see Ledlow
et al.\ 2003; Arnaud 2005, and references therein), which is
impossible to explain if clusters are fully relaxed structures. All
these observations seem to depict a process of formation for
clusters of galaxies which is more complex than what was previously
assumed (West, Jones, \& Forman 1995; Burns 1998; Forman et al.\
2003). Within this new context, a more thorough study of BCM
peculiar velocities seems warranted.

Most studies published so far on BCM peculiar velocities are based
on samples of limited size (not more than 30) or on samples that are
biased towards special types of galaxies, like D or cDs. In order to
alleviate these limitations we have built a new sample of BCMs,
based on Bautz-Morgan and Rood-Sastry types, so as to select all the
Abell clusters harboring a dominant galaxy. Our sample is
sufficiently large and general to allow a statistically significant
comparison of the BCM peculiar velocities and their morphological
types with the global properties of their clusters, and test the
models of their formation.

The organization of this paper is the following. In Section~2, we
describe our selection method for BCM recognition and present our
sample. In Section~3, we study the relation between the peculiar
velocities of the BCMs and the global characteristics of their
clusters, concentrating on different subsamples with well defined
properties. In Section~4, we discuss the dynamical implications of
our observations and introduce the merging-groups scenario for the
BCMs formation in cluster. A brief summary and our conclusions are
presented in Section~5.

\section{Selection of the sample and properties of the BCMs}

\subsection{Identification of the BCMs}

From the Abell, Corwin, \& Olowin (1989, ACO) catalog of clusters of
galaxies, including supplementary (S) clusters, we selected clusters
which have the highest probability to possess one dominant galaxy.
To this end, we chose all clusters that have a Bautz-Morgan type I
or I-II (Bautz \& Morgan 1970; hereafter denoted as BM type). Note
that the definition of BM types used by ACO is not based on the
morphology, as was the original intention of Bautz \& Morgan (1970).
As a second criterion, we also selected clusters with a Rood-Sastry
type cD (Rood \& Sastry 1971; hereafter denoted as RS type). For the
southern and supplementary samples, which do not have an RS
classification, we selected the clusters based on the comments in
Tables 7A, B, and C of the ACO paper, that suggest the presence of a
galaxy of cD type or one with a ``corona''. The application of these
criteria resulted in an initial sample of 1207 clusters.

To identify the BCMs in our sample, we used R-band images of the
Second-Epoch Digitized Sky survey (DSS2, see e.g.\
archive.stsci.edu/dss) centered on the cluster positions published
by ACO, or, occasionally, on more precise positions found in the
literature. The sizes of the images were chosen according to the
cluster redshifts ($z$) to cover a circle of typically, and at
least, half an Abell radius (Abell 1958: $R_A = 1.7'/z = 1.5\,h^{-1}
$Mpc, where H$_0 = 100\,h$\,km\,$s^{-1}$\,Mpc$^{-1}$ is the Hubble
constant). For example, we used images of $40'\times40'$ for
clusters with $z<0.045$, $30'\times30'$ for $0.045\le z<0.06$ and
$20'\times20'$ for $z\ge0.06$. Cluster redshifts were taken from the
most recent upgrade of the compilation by Andernach \& Tago (see
Andernach et al.\ 2005 for a description), with redshifts for
$\sim$110,000 individual cluster members in $\sim$3700 ACO clusters
(as of Dec.\ 2007). If a cluster had no spectroscopic redshift, we
used photometric estimates based on the work of Peacock \& West
(1992), as kindly provided by M.~West.

Visual inspection of these images allowed the identification of
usually one, sometimes two or three, and rarely four BCM candidates.
We then used the ``NASA/IPAC Extragalactic Database'' (NED,
nedwww.ipac.caltech.edu), HyperLEDA (leda.univ-lyon1.fr) or the
compilation by Andernach \& Tago to retrieve available redshifts and
other basic parameters like names and magnitudes for these
candidates. The list of these BCM candidates is given in Table~1,
with further properties compiled in Table~2.

In many clusters the first obvious candidate turned out to be a
luminous foreground or background galaxy, as judged on its redshift
compared to the cluster mean or photometric estimate. In general,
any candidate whose radial velocity differed by more than
2500\,km\,s$^{-1}$ from the cluster mean, was rejected as a BCM. In
a few cases, a somewhat smaller difference was applied after
detailed analysis of the cluster member velocity distribution. Note
that we also excluded as BCM a few additional galaxies without
redshift, like, for example, MCG$+10-17-046$, a 16$^m$ galaxy
located to the north of A1351, which is obviously far too luminous
to be member of the assumed associated z=0.32 cluster. As a
reference for future work in the field, the list of the 238 rejected
BCMs associated to 192 clusters is given in Table~3.

Whenever a BCM candidate for one cluster was found in the foreground
or background of the cluster, the cluster image was reinspected for
further possible candidates, until either a member galaxy was found
and included in Table~1, or else the cluster itself was discarded as
a whole from the sample, for its lack of a dominant galaxy. The
latter occurred for 38 clusters, which we list in Table~4.

Frequently a cluster was found to be a superposition of different
clusters along the line of sight, as judged from available redshifts
in the cluster region. These ``redshift components'' of the same
original Abell cluster are distinguished in Table~1 with capital
letters appended to the cluster number (A through E, according to
increasing redshift). Occasionally we were able to identify a BCM in
more than one such component of the same Abell cluster. We shall
refer to these clusters as the ``superposed clusters''.

In 165 cases we accepted more than one candidate as BCMs in the same
redshift component of a cluster. In Table~1, these are identified by
small letters appended to their name (a, b, etc., in order of
decreasing brightness). This apparent multiplicity may have various
reasons: either there are two real dominant galaxies, or the lack of
dominance of the brightest galaxies motivated us to include more
than one galaxy, or simply the difficulty to distinguish small
magnitude differences between the candidates by eye. In 33 cases we
accepted a third candidate as BCM, marked as ``c'', and in 6 cases a
fourth one, marked as ``d''.

The cases of foreground (background) galaxies and superposed
clusters may have affected the BM types and possibly the richness
class of the assumed associated clusters, since these
classifications were performed mostly before the cluster and BCM
redshifts were known (or in other cases this redshift information
was simply not taken into account). As already discussed by Leir \&
van\,den\,Bergh (1977) this may also have affected other BCM studies
in the past. Apart from repeating the original uncertainty flags on
the BM type by ACO in Table~1 (``:'' and ``?'') , we have put the BM
types within parenthesis whenever the cluster has a rejected BCM
listed in Table~3. In our statistical analysis we deliberately
exclude these when necessary.

\subsection{Description of Samples: Tables 1 to 4}

Tables~1 and 2 contain data and properties for 1426 BCMs in 1221
redshift components (see explanation above) of 1169 distinct Abell
clusters. This is, by far, the largest compilation of BCMs
associated with Abell clusters published up to now.

In columns~2 and 3 of Table~1, we give the RA and DEC of the BCMs.
The position was measured by fitting a bidimensional Gaussian on its
image, using the NRAO program
FITSview\footnote{http://www.nrao.edu/software/fitsview; The (USA)
National Radio Astronomy Observatory (NRAO) is operated by
Associated Universities, Inc. and is a Facility of the (USA)
National Science Foundation.}. Except for the very nearby BCMs which
extend over several arcminutes, they have a typical uncertainty of
$0.5''$.

Note that since the general rms uncertainty of the cluster center
positions published by ACO is known to be $\sim3'$, and since good
X-ray positions are available only for a small fraction of clusters
in our sample, we refrained from calculating angular offsets of the
BCM positions with respect to their cluster centers.

In Table~1, we also list the properties of the host clusters of the
BCMs, as found in the upgraded compilation of Abell cluster
redshifts by Andernach \& Tago.  These properties are: in column~4,
the Abell richness,  in column~5, the BM type of the cluster
(converting roman numbers I, I-II, ... III, to 1, 2, ... 5), and in
column~6, its RS type from Struble \& Rood (1987) and Struble \&
Ftaclas (1994).  For southern clusters without RS type, we indicate
whether the notes in Tables~7A, B, or C of the ACO paper suggest the
presence of a galaxy of cD type (listed as ``NcD'') or one with a
``corona'' (listed as ``Ncor''). The heliocentric cluster mean
velocity, $v_{cl}$, or a photometric estimate if appended with the
letter ``e'', follows in column~7, as taken from the updated version
of the ACO redshift compilation by Andernach \& Tago. An appended
letter ``n'' on this value means that these velocities were taken
directly from NED or the literature; a colon marks an uncertain
redshift, and an asterisk indicates a value differing by more than a
factor two from the photometric redshift estimate. The number of
galaxies, $N_z$, used to determine the redshift of the cluster is
given in column~8, followed in column~9 by the velocity dispersion
of the galaxies in the cluster, $\sigma_{cl}$, determined using the
method developed by Danese et al.\ (1980). In column~10 we give the
heliocentric velocity for 1032 (73\%) of the BCMs, usually from NED.
An asterisk appended to this last velocity indicates that it comes
from Andernach \& Tago's cluster redshift compilation, or
occasionally from HyperLEDA.

It is interesting to note that among the clusters without a radial
velocity for the BCM in Table~1, there are 60 clusters with
$N_z\ge5$, of which 22 have $N_z\ge10$. We have undertaken a
spectroscopic observing program in both the northern and southern
hemispheres to obtain velocities for most of these galaxies (these
velocities are not included in the present paper).

In Table~2 we give individual properties of the BCMs listed in
Table~1. Column~2 gives the peculiar velocity of the BCM with
respect to the mean velocity of its host cluster. This value is
calculated using the following relation:

\begin{equation}
v_{pec}=\frac{v_{BCM}-v_{cl}}{(1+z_{cl})}
\end{equation}

\noindent where $v_{BCM}$ is the heliocentric velocity of the BCM,
$v_{cl}$ and $z_{cl}$ is the heliocentric mean velocity and redshift
of its host cluster. The term $(1+z_{cl})^{-1}$ is a cosmological
correction (Danese et al.\ 1980). Note that we quote $v_{pec}$ only
if $N_z\ge10$, i.e.\ when $v_{cl}$ is based on at least ten cluster
members with measured redshift.

For comparison purposes, we also give in column~3 the relative
peculiar velocity of the BCM, which is the peculiar velocity in
units of the cluster velocity dispersion, $\sigma_{cl}$. For two
clusters with $N_z\ge10$ (A0136 and A3088B) no $\sigma_{cl}$ is
available, since neither $\sigma_{cl}$ nor individual galaxy
redshifts were published. Column~4 gives the BCM morphological type,
as determined by us (see explanations below), followed in column~5
by the galaxy identification from 2MASS, and other names as found in
the literature in column~6. The latter two column entries were both
retrieved from NED.

The morphological types of the BCMs in column~4 of Table~2 were
determined from visual inspection of the DSS images. Following
Morgan (1958), we define a D~galaxy as an elliptical galaxy (E) with
an extended, low-surface-brightness envelope, where the envelope is
at least two times larger than the high-surface-brightness central
region of the galaxy. Further to this first definition, we define a
cD galaxy as a giant D galaxy (Matthews et al.\ 1964). By ``giant''
we mean the galaxy is apparently the largest or most extended galaxy
of the cluster. When it is not possible to distinguish clearly
between these cases using these definitions, we classify the galaxy
as E/D (i.e.\ a possible D galaxy) or D/cD (i.e.\ a possible cD
galaxy). The only further classification we used is the spiral type
(S). Again, the class E/S (possibly a spiral) was used when it was not
possible to distinguish between the two. Except for very few cases,
it was impossible to be more explicit on our classification of the
spirals in our sample. Consequently, the type S is used in Table~2
to describe either an S0, Sa or Sb galaxy.

In many cases, the above morphological types seemed insufficient to
describe the galaxies. This is particularly true for cD galaxies.
For example, in Table~2 we classified the few cases of possible
``dumbbell'' cD galaxies as ``cD db''. Frequently also we were able
to distinguish some structures within the envelope of the cD. While
these structures could be mere superpositions, they suggest possible
evidence for mergers, explaining our label as ``cD\,m'' in Table~2.
Finally, we found many systems formed by two or more, apparently
elliptical galaxies. Without prejudice on the nature of these
objects, we marked them as possibly interacting (``int'') in
Table~2. A high level of uncertainty on our classification is
indicated by an interrogation sign. Note that only five galaxies
($\sim0.3\%$) in our sample could not be classified at all.

In Table~3, columns 2 and 3 give the positions (also measured by us)
of the discarded BCMs, followed in columns 4 and 5 by the
heliocentric mean velocities of the presumed associated clusters and
actual redshifts of the discarded BCMs. We also give in column~6 the
names of these galaxies as found in NED.

\subsection{Distribution of BCM Morphology and Adoption of our Main Sample}

We separate our sample into two statistical groups. In the Primary
sample we put our ``primary'' BCMs. This includes the BCMs
associated with the components of superposed clusters and those
marked as ``a'' for the multiple cases. In the Secondary sample we
put the ``secondary'' BCMs which includes 211 galaxies marked as b,
c or d in Table~1. In these two subsamples, we then eliminate all
the BCMs with an uncertain morphological classification, i.e.\ all
galaxies for which the morphological type in Table~2 contains a
question mark. This leaves 1125 galaxies in the Primary sample and
193 in the Secondary sample.

The distribution of morphological types of the galaxies in the two
subsamples are shown in Figure~1. The dominant morphological type in
the Primary sample is cD (35\% of the sample). Together with the
BCMs classified as D/cD they represent 48\% of the whole sample. The
second most frequent type is E (19\%). Adding to these the D and E/D
types, they form 44\% of the whole sample.  In the Primary sample,
therefore, very few BCMs (8\%) are spiral-like (i.e.\ of type E/S or
S).

In the Secondary sample we see a shift towards later morphological
types compared to the Primary sample. The dominant morphology in
this group is E (44\%), followed by the E/D and D which added
together form (73\%) of the whole sample. Even the S type (10\%),
which together with the E/S form 22\% (there is no D/cD in the
statistical group S), is more abundant than the cD (5\%).

The distribution of the morphologies in the Primary sample is
consistent with what we expect for BCMs: galaxy clusters are systems
dominated by early-type galaxies, many of which (the D and cDs) are
unique to clusters. The shift towards later morphological types in
the Secondary sample is consistent with the expectation that
second-, third-, and fourth-brightest galaxies in clusters are less
``evolved'' morphologically as compared to the dominant BCM.

Taken at face value, this result seems to support our choice of
BCMs. For our statistical analysis, the BCMs in the Primary sample
will form our main sample.

Before looking into the relation between morphology and peculiar
velocity, we note one further interesting characteristic of their
morphological distribution: not all BCMs are cDs. This is somewhat
surprising considering the bias introduced by our selection
criteria. Indeed, we would have expected clusters with a dominant
galaxy to host mostly cDs. Considering the size of our sample one
can interpret this result in two ways: 1) assuming cDs are part of
the normal evolution of E galaxies, then most clusters in our sample
did not reach a stage of evolution sufficient to produce them; 2)
assuming cDs are a special phenomenon, then not all the clusters in
our sample possess the conditions required to form them.

\section{Analysis}

\subsection{Peculiar Velocities of the BCMs}

The main concern of our study is the peculiar velocity, $v_{pec}$,
of BCMs in clusters. As mentioned previously, the peculiar
velocities in our sample were calculated only when the number of
galaxies with redshifts measured to determine the dynamical
characteristics of the cluster, $N_z$, is greater or equal to 10.
This reduces our main sample to 452 BCMs. The distribution of
relative peculiar velocities is presented in Figure~2. The mean,
median and percentile values for $|v_{pec}|/{\sigma}_{cl}$ for the
whole sample, as well as for the different morphological types of
BCMs, are reported in Table~5. Based on these statistics, we
conclude that 50\% of the BCMs in our sample have a peculiar
velocity higher than 32\% the velocity dispersion of their cluster.

Our analysis confirms the findings obtained by different authors: an
important number of BCMs in clusters have significant peculiar
velocities. Before us, these findings were based on rather small and
incomplete samples, which were prone to statistical fluctuations: 22
clusters in Malumuth et al. (1992), 31 in Zabludoff et al. (1993),
25 in Bird (1994) and 25 in Oegerle \& Hill (2001). The generality
of our result (452 clusters) clearly establishes this phenomenon as
a common trait of clusters of galaxies harboring a dominant galaxy,
and not a special feature of clusters with D- or cD-like BCMs.

\subsection{Robustness and Errors of Peculiar Velocity}

To test the robustness of our definition of peculiar velocity, we
look for a possible dependence on the number of galaxies used to
calculate the cluster velocity dispersion. To do so, we separate our
sample into four different statistical groups: N10 (452 BCMs)
contains clusters with $N_z\ge10$ measured redshifts, N30 (243 BCMs)
contains clusters with $N_z\ge30$, N50 (151 BCMs) those with
$N_z\ge50$, and N100 those with $N_z\ge100$ (64 BCMs). The
box-whisker plots for the distribution of BCM peculiar velocities in
the different subsamples are presented in Figure~3. Note that we use
the absolute peculiar velocity instead of the relative one to avoid
the obvious problem of the increase of velocity dispersion with
richness (clusters with more than 100 galaxies measured are among
the richest and have consequently higher velocity dispersions, which
reduces, somewhat artificially, the relative peculiar velocities of
their BCMs). The median or mean peculiar velocities do not decrease
significantly with increasing number of redshifts, i.e.\ when going
from subsamples N10 to N100.

In order to establish the statistical significance of this result,
we perform a non-parametric Kruskal-Wallis (K-W) test on these
subsamples. A non-parametric test is used because the distributions
are not Gaussian (as verified using three different tests:
Kolmogorov-Smirnov, D'Agostino \& Pearson and Shapiro-Wilk normality
tests). The probability, P, calculated by the K-W test is the
probability that random sampling from populations with similar
distributions produces a sum of ranks as far apart as observed. All
the non-parametric tests used in our analysis are done at a level of
significance of 95\%, which is standard for such tests. A small
probability (P $<0.05$), suggests that the samples are unlikely to
be drawn from the same population.

The K-W test applied to our subsamples detects no significant
difference between the distributions (P~$= 0.9126$). We conclude
that our finding of a high fraction of high peculiar velocity BCMs
is independent of the number of galaxies used to calculate the
cluster velocity dispersion.

It is easy to show that our result neither depends on the errors in
the BCM radial velocity nor on those of the cluster mean, nor on
those of the cluster velocity dispersion. A typical value for the
error on the radial velocity of the BCMs is $\Delta
v_{BCM}=60$\,km\,s$^{-1}$. For the error of the cluster mean
velocity we use the standard deviation of the mean, which is $\Delta
v_{cl} = \sigma_{cl}/\sqrt{N_z}$. The error of the absolute BCM
peculiar velocity is then $\Delta v_{pec} = \sqrt {\sigma_{cl}^2/N_z
+ (\Delta v_{BCM})^2}$. For our sample of 452 BCMs we find that 41\%
have peculiar velocities of more than twice this error, and 25\% of
them with more than three times their error. Assuming a very
conservative value of $\Delta v_{BCM} = 100$\, km\,s$^{-1}$, we
obtain 33\% of BCMs  with $v_{pec} \ge 2 \Delta v_{pec}$, and 19\%
with $v_{pec} \ge 3 \Delta v_{pec}$. To estimate the error in the
{\it relative} peculiar velocity we assumed an error in the cluster
velocity dispersion of $\Delta\sigma_{cl}$ = 100\,km\,s$^{-1}$. This
is a reasonable estimate as this error depends more on the median
error of the individual galaxy velocities than on the total number
of cluster members (see Danese et al.\ 1980; Adami et al, 1998;
Fadda et al., 1996 or De~Propris et al. 2002). The error of the
relative peculiar velocity is then $\Delta(v_{pec}/\sigma_{cl}) =
\sqrt {(\Delta v_{pec})^2 + (v_{pec} \Delta\sigma_{cl} /
\sigma_{cl})^2} /\sigma_{cl} $. For the very conservative value of
$\Delta v_{BCM}$ = 100\,km\,s$^{-1}$ we find that of our 452 BCM
candidates 31\% have $v_{pec}/\sigma_{cl} \ge 2 \Delta
(v_{pec}/\sigma_{cl})$, and 13\% have $v_{pec}/\sigma_{cl} \ge 3
\Delta (v_{pec}/\sigma_{cl})$. We conclude, therefore, that the
uncertainties in BCM and cluster mean radial velocities, as well as
those in the cluster velocity dispersion, are not the cause for a
significant fraction of high peculiar velocity BCMs.

\subsection{Relations between Peculiar Velocity, BCM Morphology,
Cluster Richness and BM Type}

Our large sample allows us to go further in our analysis by
searching for a possible relation of the peculiar velocity with the
BCM morphology, the cluster richness and the BM type.

To check for a relation between BCM relative peculiar velocity and
BCM morphology, we separate our sample in three subsamples: the
cD~sample is obtained by merging the D/cD galaxies with the cD, the
D, E/D and E are grouped into the E~sample, and all the other
galaxies (E/S and S type) are grouped into the S~sample. The
distributions for the three subsamples are shown in Figure~4. The E
galaxies trace a more homogeneous distribution than the cDs, who
seem to aggregate at lower relative peculiar velocities.

Our statistics for the three subsamples are reported in the
rightmost columns of Table~5. We clearly distinguish a tendency for
galaxies in the cD subsample to have lower relative peculiar
velocities: the median $|v_{pec}|/\sigma_{cl}$ is 0.27 for the cD
subsample, compared to 0.45 for the E~sample. Curiously, the median
decreases again to 0.38 in the S~sample. However, this last sample
is quite small and prone to larger statistical uncertainties (as
suggested in Table~5 by the relatively large standard error on the
mean).

In order to verify the statistical significance of the differences,
we perform a K-W test on these subsamples, since, once again, the
distributions are not Gaussian. The probability for our three
subsamples to be drawn from the same population is only P $=0.0002$,
which is extremely significant. A post-test (Dunn's multiple
comparison test) allocates the difference between the cD and E
subsamples. No significant differences are encountered between the E
and S samples, or between the cD and S samples. Application of a
different non-parametric test, the Mann-Whitney (M-W) test, which
compares only two samples at a time, confirms these results.

We conclude that the relative peculiar velocity of a BCM depends
very strongly on its morphological type. In general, therefore,
cD-type BCMs have lower relative peculiar velocities than D, E/D or
E together.

To investigate the effect of cluster richness we separate our sample
into three subsamples. The subsamples identified as R0 and R1
contain clusters with respective Abell richness classes R $=0$ and R
$=1$. Those with a richness class~2 or more were grouped into the
subsample R2+. Since the richness class of the superposed clusters
(i.e.\ those with component letters A, B, ...) is questionable, we
excluded them from our statistics. That leaves us with 276 BCMs. The
distributions for the three subsamples are shown in Figure~5a. The
medians, percentiles and means for $|v_{pec}|/{\sigma}_{cl}$, for
BCMs in clusters of different richness and BM type, are reported in
Table~6.

Comparing the distributions in Figure~5a, and the medians in
Table~6, we do not see any significant changes in the BCM relative
peculiar velocity distribution when passing from the richness
samples R0 through R1 to R2+. This is not only confirmed by the high
P value of a K-W test: P~$=0.2026$, but also by individual M-W
tests. We conclude that, in general, the relative peculiar velocity
of a BCM does not depend on the richness of its host cluster.

To check for a relation between BCM relative peculiar velocity and
its host cluster BM~type, we divide our sample into two subsamples:
clusters with BM~type~I form the subsample BM1 and those with a
BM~type I-II form the subsample BM2. Clusters with a BM~type II or
later were included in our main sample only because they contain a
cD, or were suspected to contain one. They are thus not fully
represented in our sample and are not considered in the present
comparison. The superposed clusters are also excluded, since their
BM~types must be considered as uncertain. This leaves us with 189
BCMs. The distributions for the two subsamples are shown in
Figure~5b and the statistical results are also reported in Table~6.

We observe a significant variation in the relative peculiar
velocities of the BCMs, passing from the BM1 clusters to the BM2
clusters: the median increases from 0.26 to 0.40. A value of
P~$=0.0017$ from a M-W test implies that the difference is very
significant. This suggests that the dominant nature of a BCM in its
cluster favors lower relative peculiar velocities.

Note that this last result is consistent with the relation with
morphology, since we expect cD galaxies to be the dominant galaxies
in their host clusters. We suspect, therefore, that the strong
relation encountered between the relative peculiar velocity and
morphology is the main cause of the relation found with the BM type.
This interpretation will be checked in the following section.

\subsection{Relations between BCM Morphology, Cluster Richness and BM Types}

To better understand the nature of the relations (or absence
thereof) found in the previous section, it is important to establish
which connections exist between the different parameters studied.

As judged from Figure~6a, there is a definite increase in the
fraction of cD-type BCMs in clusters of earlier BM type: 46\% in BM1
clusters compared to 22\% in BM2 clusters. There are also many more
BCMs of types E/D and E in clusters of later BM type: 48\% in BM2
clusters compared to 27\% in BM1 clusters. This result is expected,
considering the definition of cDs as dominant galaxies.

We extracted the cD-type galaxies in clusters with BM types I and
I-II and performed a M-W test on the medians of their relative
peculiar velocity.  No significant difference (P~$=0.1239$) was
found. In other words, cD galaxies have similarly low relative
peculiar velocities, independent of the BM type of their host
cluster. This supports our interpretation that the relation between
the BCM relative peculiar velocity and its host cluster BM type is
due to the fact that BCMs in BM\,I clusters are mostly cDs.

In Figure~6b, we compare the distributions of the morphologies of
the BCMs in the three richness subsamples, as defined before. There
is a definite rise in the fraction of the cD morphology for the BCM
in richer clusters. The fraction of cDs increases from 25\% in R0
clusters to 48\% and 57\% in R1 and R2+ clusters, respectively.
Consequently, there are slightly more BCMs of later types (later
than D) in low-richness clusters: 55\% in R0 clusters, compared to
33\% in R1 and 23\% in R2+ clusters.

Therefore, although we find no direct relation between the relative
peculiar velocity of the BCM and the host cluster's richness we do
find a trend for richer clusters to harbor cDs, which have lower
relative peculiar velocities.

\subsection{Relation with Cluster Velocity Dispersion and mass}

The cluster velocity dispersion is usually taken as a proxy for the
total mass of the system. This interpretation is founded on the
assumption that clusters are dynamically relaxed and follow the
virial theorem. If the distribution of the luminous mass follows
that of the total mass, then more massive clusters must also be
richer in galaxies. Consequently, we expect the richness of a
cluster to increase with the velocity dispersion. But what about the
peculiar velocity? We have just seen that there is no relation
between richness and the relative peculiar velocity, and this is
despite the fact that the frequency of cDs is higher in richer
clusters.

To explore this point we examine how the velocity dispersion varies
as a function of the other parameters in our study. Box-whisker plots for
the velocity dispersion of galaxies in cluster subsamples separated by
Abell richness and BCM morphological types (using the same regroupment
as before) are presented in Figure~7.

In Figure~7a, we distinguish a very strong increment of the velocity
dispersion with the richness of the cluster, which is also confirmed
by the statistics in Table~7, where we report the median,
percentiles and mean cluster velocity dispersion for clusters of
different richness. The K-W test detects extremely significant
differences (P~$< 0.0001$). The post-test allocates the most
significant differences between the R0 and R1 and between the R0 and
R2+ subsamples. No significant difference is detected between the R1
and R2+ subsamples by the post-test. However, the result of a M-W
test between the R1 and R2+ subsamples find a significant difference
(P~$=0.0129$).

If we take the velocity dispersion as a proxy for the cluster mass,
then massive clusters are richer in galaxies.

In Figure~7b, we distinguish a definite increase of the host cluster
velocity dispersion when passing from the BCM type S through E and
cD subsamples. The statistics reported in Table~7 confirm this
observation, where the rightmost three columns give the mean,
percentiles, and median velocity dispersion of clusters with
different morphological types of BCMs.  The K-W test detects
extremely significant differences (P~$< 0.0001$). The post-test
identifies extremely significant differences between the cD and E
and very significant difference between the cD and S samples. The
post-test does not detect a difference between the E and S samples.
The high P value (P~$=0.2524$) for a M-W test between the E and S
subsamples confirms this last result.

Looking at Figure~7, and considering the statistical tests, we have
to conclude that the relation between cluster richness and cD
galaxies is consistent with the following interpretation: cDs are
more common in rich clusters probably because rich clusters are
generally more massive.

In Figure~8, we compare the velocity dispersion and the absolute
value of the BCM peculiar velocity, $|v_{pec}|$. For this test we
consider only the cD~subsample and the E~subsample. In Figure~8a we
see a very weak trend in the cD subsample, suggesting that the
peculiar velocity increases with the velocity dispersion. The trend
is more obvious in the E subsample (Figure~8b). Comparable trends
were previously observed by Malumuth et al. (1992) and Bird (1994).
Contrary to these authors, however, we do not confirm a similar
correlation with the richness of the clusters. In both graphs, we
distinguish between the different richness classes. We see no
particular difference in these trends for the different richness
classes.

To verify if the trends we observe are statistically significant, we
perform two Spearman correlation tests. The tests yield a
correlation coefficient r~$=0.27$ for the cD subsample, and
r~$=0.41$ for the E~subsample, both with a probability P~$<0.0001$,
consistent with extremely significant positive correlations. This
implies that, in general, the BCM peculiar velocity rises as the
cluster velocity dispersion increases.

In Figure~9, we show the box-whiskers plots for the peculiar
velocities as found in the cD and E subsamples separated by richness
classes. We find no differences in the peculiar velocity between the
richness classes. The statistics for these two subsamples are
reported in Table~8. The K-W tests detect no differences (P~$=
0.3265$ for the cD subsample and P~$= 0.9680$ in the E subsample)
for the medians in the subsamples separated by richness classes.
This result confirms that the cluster richness plays no role in the
correlations found in Figure~8 and in the general frequency
distribution shown in Figure~4.

\section{Discussion}

Our analysis confirms the findings of previous authors working in
the field (Beers \& Geller 1983; Tonry 1985b; Malumuth et al.\ 1992;
Zabludoff et al.\ 1993; Bird 1994; Oegerle \& Hill 2001; Pimbblet et
al.\ 2006): most BCMs are not at rest at the center of their host
cluster's potential well. The large size and completeness of our
sample eliminate any doubts on the physical reality and generality
of this phenomenon. Our analysis also shows that this is a common
trait of clusters of galaxies harboring a dominant galaxy and not a
special feature related to particular systems, like clusters hosting
a D or cD galaxy.

There is no easy way out of this situation. Assuming, for example,
that the BCMs are really at rest at the dynamical center of their
clusters would raise the peculiar velocity of the other galaxies,
putting them at higher energy levels in the potential well of their
clusters. To explain the observations assuming dynamical equilibrium
would increase the amount of dark matter to possibly unacceptably
large values (for instance, in terms of $M/L$, see Tonry 1985a for
an explanation).

The fact, also, that the peculiar velocity, a dynamical parameter
related to the cluster, is strongly correlated with the morphology
of the BCMs, seems to suggest a strong connection between the
formation of a cluster and its BCM. For example, assuming BCMs form
by the mergers of smaller mass elements, we would naturally expect
massive galaxies (D and cDs) to be more frequent in richer clusters,
which is consistent with our analysis. This is because the number of
mergers, or the masses of the merging components, is expected to
grow with the mass in the clusters. However, richer clusters also
have higher velocity dispersions, which reduce the efficiency of
mergers (Tonry 1985a; Mihos 2004). Unfortunately, the current status
of simulations of large-scale structures formed by cold dark matter
is not of much help. These models do not include the physics of
galaxy formation and the best simulations to date place, more or
less artificially, the BCMs at the center of the halos (e.g.\ Taylor
\& Babul 2004; Springel et al.\ 2005, or De\,Lucia \& Blaizot 2007),
predicting zero peculiar velocities.

\subsection{Explaining the BCM peculiar velocities}

Let us re-examine the present paradigm of structure formation to see
how it may be adapted to fit our observation. According to the
model, 90\% of the mass of a cluster is in the form of non-baryonic
dark matter. This follows directly from the standard cosmological
scenario, in which dark matter perturbations are free to grow as
soon as they enter the particle horizon, while baryonic matter can
do so only after it decouples from radiation. In fact, this is the
strongest argument in favor of the existence of dark matter, since
structures dominated by non-baryonic dark matter could grow to
significant masses without producing anisotropies in the microwave
background in excess of what is observed. For CDM cosmology the
first structures to form after recombination ($z = 1000$) have
typical masses of the order $10^5$ M$_\odot$ (e.g. Coles \& Lucchin
1997, or any good book on cosmology and structure formation).

After decoupling the physics becomes non-linear and numerical
simulations are necessary (see Davis et al. 1985 and references
therein). Reviews of this subject can be found in Primack (1999),
Arnaud (2005), or Loeb (2008). To summarize, within the CDM
paradigm, structure formation follows a bottom-up scenario, where
high-mass halos gradually form from the mergers of smaller mass
ones. The question is how to include consistently the formation of
the BCMs and their peculiar velocities into this model?

Numerous simulations show that in any self-gravitating system, the
most massive galaxies are expected to lose energy through dynamical
friction to the less massive bodies and to spiral towards the bottom
of the potential well (White 1976; Merritt 1983; Tonry 1985a;
Malumuth 1992). Following Tonry (1985a), the dynamical friction
decay of velocity of as galaxy with path length $x$ is given by:
\begin{equation}
\frac{{dv}} {{dx}} =  - C\frac{{M\rho }} {{v^3 }}g\left( v \right)
\end{equation}
where $M$ is the mass of the galaxy, $v$ its velocity, $\rho$ the
density of the background medium and $g(v)$ a function that depends
on the distribution of velocities of the background particles. For
an isothermal distribution of velocities with dispersion $\sigma$
the equation takes the form:
\begin{equation}
\frac{{dv}} {{dx}} =  - C\frac{{M\rho }} {{\sigma ^3 }}\frac{1}
{{\alpha  + \left( {{v \mathord{\left/
 {\vphantom {v \sigma }} \right.
 \kern-\nulldelimiterspace} \sigma }} \right)^3 }}
\end{equation}
where $\alpha$ is a geometric constant. Dynamical friction increases
with the mass of the galaxy and the density of the background
particles, while it decreases with the velocity dispersion of the
background particles. This seems consistent with our observations
(assuming cDs are more massive than E galaxies).

On the other hand, what seems difficult to understand is why after a
Hubble time, most BCMs are not at rest at the bottom of the
potential well of their clusters. Indeed, slightly less than a third
(29\%) of the BCMs in our sample may be consistent with zero
peculiar velocity. According to Malumuth et al. (1992), this
phenomenon should be viewed as evidence of a relatively recent
formation. This is because, in hierarchical structure formation
models, the richest, most massive systems must have undergone the
most recent merger events. Based on this interpretation, Malumuth et
al. (1992) proposed that high peculiar velocity BCMs must occur only
in rich, high velocity dispersion clusters. This is not confirmed by
our analysis. Although we do observe a positive correlation between
the peculiar velocity and velocity dispersion of the clusters, we do
not distinguish the trend expected with the richness. Neither can we
find, according to this interpretation, a natural explanation why
the correlation is stronger in the E subsample than in the cD
subsample. In general, we do not observe any specific dynamical
characteristic that allows to distinguish the clusters with low
peculiar velocity BCMs from those hosting BCMs with high peculiar
velocities.

Can we explain the peculiar velocities using a special form of dark
matter halo? In his article Tonry (1985a) explains that the matter
density, $\rho$, of the background matter must play a major role.
For example, when a cluster has a radial density profile that is
cuspy, like for an isothermal sphere (with density as function of
radius: $\rho (r) \propto r^{-2}$), the dynamical friction in the
center of the cluster is stronger than when the density falls less
rapidly with radius. Consequently, if the global halos of clusters
have such a shallow central density profile the orbits of massive
galaxies may take longer to decay. The Navarro, Frenk \& White (NFW)
halo model (density: $\rho (r) \propto (r/r_s)^{-1} (1+r/r_s)^{-2}$,
where $r_s$ is a scale radius) seems to show such a property.
Therefore, if dark matter in clusters follows originally such a
distribution, even after a Hubble time BCMs may not have had
sufficient time to relax dynamically, explaining their non-zero
peculiar velocities. However, it may be that the free parameters in
the NFW model would need to vary significantly from one cluster to
another to accommodate all our observations.

Even if we consider non-zero peculiar velocities possible within the
NFW model (as an oscillation of the BCM around the center of the
potential well) there would still be one more important difficulty.
Because the intracluster gas producing the X-rays is ten times more
massive than the luminous matter in galaxies, we should not expect
this gas to follow the oscillating BCMs. However, most observations
seem to suggest just that: the large majority of BCMs are located at
the peak of the X-ray emission (Jones \& Forman 1984; Rhee \& Latour
1991; Bahcall et al.\ 1995; Bahcall 1999; Mulchaey et al.\ 2003).

As a preliminary verification, we cross-correlated the X-ray peak
positions of X-ray clusters as published by Magliocchetti \&
Br\"uggen (2007) with the positions of BCMs in a sample of Abell
clusters, even larger than the present one (article in preparation),
and found 76 clusters in common (with 46 BCMs in our present
sample). The consistency between the positions of the X-ray peaks
with the positions of the BCMs is impressive: only 8 out of 76 BCMs
show a positional offset larger than 20$''$. Of the 46 clusters in
common with Magliocchetti \& Br\"uggen (2007) and our present
sample, 41 BCMs have an offset of less than 20$''$ from the
cluster's X-ray emission peak. All these clusters have a unique BCM
in our Table~1, and 36 have a value for $v_{pec}/\sigma_{cl}$ in
Table~2. Their median $|v_{pec}/\sigma_{cl}|$ is 0.28, and there are
eight clusters with $|v_{pec}/\sigma_{cl}|>$0.6. We classified 29 of
these 36 BCMs as ``cD'' in Table~2, and these have a median
$|v_{pec}/\sigma_{cl}|$ of 0.26. Thus, a significant fraction of
high peculiar velocity BCMs persists in subsamples where the
positional coincidence between the BCMs and their X-ray peak is very
good.

In conclusion, it seems difficult to explain the peculiar velocities
of the BCMs as an oscillation component around the center of a
global potential well formed by a spherical halo of dark matter,
which is already dynamically relaxed. And this is true even if the
halo has a shallow central density, like in the NFW model. Also,
based on X-ray observations, the local potential well formed by the
BCM must also be that of the dark matter halo of the cluster. In
other words, it seems impossible to separate the dark matter halo of
the clusters from that of their BCMs (Bahcall 1999).

Another intriguing result of our analysis is the correlation between
the peculiar velocity and morphology of the BCM. In principle, the
dissipative processes involved in galaxy formation are unrelated to
the process that pulls the BCMs towards the center of the cluster.
Consequently, we would not expect the peculiar velocity of a BCM,
which depends on the latter, to be related to the morphology of the
galaxy, which depends on the former. Therefore, the fact that we do
observe a trend for cD galaxies to have smaller relative peculiar
velocities can only be explained if the dissipative processes
related to the formation of the BCMs are somehow connected to the
force that is pulling them towards the center of the potential well
of their clusters. This suggests that the peculiar velocities of the
BCM must reflect not only the formation of the BCMs within the
clusters but also the process by which the clusters formed. The two
phenomena cannot be separated.

More exotic explanations of peculiar velocities, like the effect of
gravitational redshifts (Cappi 1995; Broadhurst \& Scannapieco 2000;
Kim \& Croft 2004), can also be readily eliminated. Among the
parameters that contribute to the velocity difference between a BCM
and its cluster, the gravitational redshift component is always
positive (Kim \& Croft 2004). The effect of gravitational redshifts
would therefore skew the distribution of peculiar velocities towards
positive values, while the observed peculiar velocity distribution
is very symmetrical about zero. Thus we agree with Kim \& Croft
(2004) that there is currently no detectable evidence for
gravitational redshifts in clusters of galaxies.

\subsection{The Merging-Groups Scenario}

As mentioned in the introduction, one alternative scenario proposed
to explain BCMs like D and cD galaxies is that they actually formed
in smaller systems like compact groups of galaxies (Merritt 1985);
Bird 1994; Zabludoff \& Mulchaey 1998; Pimbblet et al.\ 2006).
Indeed, the low velocity dispersion of galaxies in compact groups
render tidal interactions and mergers of galaxies much more
efficient (Merritt 1985; Tonry 1985b; Mihos 2004; Coziol \&
Plauchu-Frayn 2007). Assuming the compact groups that formed the
BCMs were more massive than today's compact groups, then giant
elliptical D galaxies and even cDs are possible consequences.

Implicit in this hypothesis, clusters must then build by the fusion
of many such groups (Ellingson 2003; Mihos 2004; Andernach \& Coziol
2007). What would then be the main condition to observe peculiar
velocities for the BCMs in clusters? It seems that the only way to
reproduce this phenomenon according to this hypothesis is to assume
clusters are still in an unrelaxed dynamical state. That is, the
BCMs still possess some of the dynamical properties of the groups in
which they formed, which translates into non-zero peculiar
velocities (Malumuth 1992). Consequently, the BCMs are not at the
centers of the global potential wells of their clusters, but rather
at the bottom of {\it local} potential wells (Beers \& Geller 1983;
Oegerle \& Hill 2001), which would be the potential wells of their
groups.

Taken at face value, the merging-groups hypothesis seems capable to
explain the peculiar velocity of the BCMs, although we have still to
verify if this hypothesis is consistent with our observations. This
may be difficult to check because the dynamical behavior of an
unrelaxed transient system, implied by this scenario, is more
complex to describe and to follow up than that of a relaxed
structure. One cannot apply the virial theorem or even assume a
simple form of density distribution and potential, and surely cannot
predict its behavior based on a simple analytical dynamical theory.
In the absence of these tools we can only offer a qualitative
evaluation, using general dynamical arguments.

We have found, for example, that the relative peculiar velocity is
smaller for BCMs of type cD, compared to any other morphological
type. This seems reasonably easy to understand. The fact that cDs
are the dominant galaxies in their clusters suggests they formed in
the most massive groups. These groups would necessarily constitute
an important fraction of the mass of their clusters, explaining the
trend towards lower relative peculiar velocities. However, of the
29\% BCMs with a peculiar velocity consistent with zero (within the
observational errors), only 36\% are cDs. Obviously, cDs are not
restricted to these cases, because it also depends on the merger
history of the cluster: cDs would be less dominant in clusters that
formed from a large number of groups. This would be consistent with
the lack of correlation of the peculiar velocity with richness and
its increase with cluster velocity dispersion.

On the other hand, we have found that a higher richness favors the
formation of cDs. As we stated earlier, the fact that cDs are the
dominant galaxies in their clusters and have lower relative peculiar
velocities suggest these galaxies formed in the most massive groups
that merged to form clusters. Massive groups most probably attract
other groups more easily, which would produce the trend with
richness.

If one thinks in terms of the density perturbation spectrum, this
last interpretation may also explain why cDs are not ubiquitous in
clusters. Being more massive, groups which formed a cD were
necessarily located in highest-density peaks. Because high-density
peaks are less frequent than lower-density ones, not all clusters
will be expected to possess a cD, consistent with our observations.

The higher the number of groups that coalesce to form a cluster, the
richer this cluster must finally be. Assuming the system is not in
equilibrium, then statistically one would expect richer clusters to
also have higher velocity dispersions. The difference here is that
we do not have relaxation, and one cannot apply the virial theorem
to deduce the mass. That is, the velocity dispersion is not a proxy
for mass. On the other hand, the increase of luminous mass with
richness is obvious.

The situation is worse if one considers the merging process that formed
the BCM in the first place. The first galaxies that merged are those that
have the smallest differences in velocities. Consequently, the
merging process itself will leave behind galaxies that have higher
differences in velocities, raising the velocity dispersion of the
group. We observe something similar in nearby compact groups of
galaxies (Coziol et al.\ 2004; Coziol \& Plauchu-Frayn 2007): the
morphologically more evolved systems (implying more mergers) are
those that have higher velocity dispersions. This similarity
suggests a continuity in the processes of galaxy formation and
evolution in different structures.

The last phenomenon we have to explain is the impressive concordance
between the X-ray peaks, produced by the intracluster gas, and the
positions of the BCMs in the cluster. The problem of the origin of
the hot gas in clusters is a very difficult one. In realistic
hydrodynamic simulations, the high efficiency necessary to form the
galaxies in clusters leaves almost no gas behind, in dramatic
contradiction with observations (Balogh et al.\ 2001). On this
matter, the merging-groups scenario may alleviate the problem.
Because groups of galaxies have lower mass than clusters, and have
consequently shallower potential wells, an inescapable conclusion
seems to be that intergalactic gas will not be especially attracted
by these systems. This is supported by observations in X-rays:
groups of galaxies occupy the fainter part of the X-ray luminosity
function or luminosity vs.\ velocity dispersion relation for groups
and clusters (Ponman et al.\ 1996; Helsdon \& Ponman 2000; Mulchaey
et al.\ 2003; Jeltema et al.\ 2006). Some new observations, though
still controversial, may even suggest groups to be less rich in gas
at higher redshifts (Spiegel et al.\ 2007).

Thus, it may be that the bulk of the intracluster gas found today in
clusters arrived there only after the clusters, and most of the
galaxies within it, were formed. By falling into the already formed
clusters the gas would have cascaded down the different
substructures that form it towards the deepest potential wells. This
is where we also expect to find the BCMs, explaining why these
galaxies are usually associated with a peak in the X-ray emission.
By cascading down the substructures of the clusters the gas would
have heated up, transforming the newly formed structures into
environments with extremely low star formation efficiency.

An early preheating phase for the intergalactic medium would
obviously help in such a scenario. Indeed, hot gas would fall even
less easily into shallow potential wells, explaining why such a huge
quantity of gas did not form galaxies. The source of energy of this
preheated gas could be related to the evolution of the first stars,
the formation of the first black holes (AGNs) or to shocks produced
by the formation of structure (Lloyd-Davies et al.\ 2000; Dav\'e et
al.\ 2001; Valageas et al.\ 2003; Dwarakanath \& Nath 2006).

This alternative scenario for the origin of the intracluster gas may
also offer a simple alternative to the problem of the contamination
of the gas by metals. Two of the mechanisms considered for this
process are ram pressure stripping of late-type spirals falling into
the clusters (Gunn \& Gott 1972), and starburst winds produced by
mergers (Schindler et al.\ 2005; Domainko et al.\ 2006; Kapferer et
al.\ 2006). In the merging-groups scenario the intense phase of
starburst activity (and possibly AGNs) is directly related to the
formation of galaxies in groups (Coziol et al.\ 1998; Coziol \&
Plauchu-Frayn 2007). This process may also have allowed a higher
level of metals to reach the intergalactic medium. This is because
the metals are more loosely bound to galaxies in a group environment
(Renzini et al.\ 1993; Metzler \& Evrard 1994; Ponman et al.\ 1996).
Contrary to groups, however, these metals would not be lost, but
swept up by the ram pressure of the intergalactic gas falling into
the newly formed clusters for the first time.

It is important to note that according to our scenario, the ram
pressure is exerted when the gas runs over the galaxies, and not the
other way around. As an analogy, one may think of falling rain
cleaning the air of its pollutants. On average, therefore, we expect
the amount of metal in the intracluster medium to be equal to the
amount encountered in all the galaxies forming the cluster
(Schindler 2003). This is because the gas must have passed through
all the galaxies on its way down the potential wells. In other
words, we expect the mixture time to be short, and possibly shorter
than in other models.

\section{Summary and conclusion}

Based on our analysis of existing BCM velocity data, we have shown
that the peculiar velocities of BCMs in clusters of galaxies cannot
be ignored. This is a general phenomenon, affecting the majority of
clusters with a dominant galaxy. We have shown that such a
phenomenon is difficult to explain within a model where the BCMs
form independently from the dark matter halo of their clusters. The
existence of a strong relation between the BCM peculiar velocity and
its morphology also points towards an intrinsic relation between the
formation of the BCM and that of its cluster.

Based on our analysis, we have found our observations to be
qualitatively consistent with a scenario where BCMs in clusters form
first in smaller mass systems comparable to compact groups (Merritt,
1985; Bird, 1994; Zabludoff \& Mulchaey, 1998; Pimbblet et al.,
2006). Implicit in this hypothesis, the formation of clusters would
have followed the merging of many such groups (Malumuth 1992;
Ellingson 2003; Mihos 2004; Adami et al.\ 2005; Andernach \& Coziol
2007; Coziol \& Plauchu-Frayn 2007). This has one immediate
consequence, which is that most clusters of galaxies harboring a
dominant galaxy are not dynamically relaxed.

Although our observation of many BCMs with large peculiar velocities
also seems in good agreement with the presence of substructures in
clusters (Bird 1994; Dressler \& Schectman 1988; Schuecker et al.\
2001; Flin \& Krywult 2006), we are not sure whether the two
phenomena are equivalent. In particular, the explanation for each of
these observations may be different. The usual interpretation of
substructures in clusters of galaxies is that they are evidence that
these systems formed recently. This seems somewhat in contradiction
with the advanced morphological stage of galaxies in clusters. These
are among the most massive and oldest (in terms of stellar
populations) galaxies in the universe.  In part, the merging-groups
scenario solves this apparent contradiction. A group environment
allows galaxies to evolve rapidly through tidal interactions and
mergers (Coziol \& Plauchu-Frayn 2007). However, one also has to
consider that the relaxation time of a cluster formed by many
groups, that is the time it takes for the energy to be redistributed
equally throughout the cluster, is probably much longer than the
typical dynamical friction time for one galaxy falling into an
isotropic potential well. In fact, the relaxation time for the
former could be much longer than the Hubble time. Consequently, it
would be possible to observe peculiar velocities, even if the
merging of groups forming the clusters started at a very early epoch
($z\sim3-4$). On the other hand, the substructures observed in
clusters today could be traces of more recent events, related to
continuous accretion of mass by the clusters, namely loose groups or
smaller groups of galaxies falling in from the field.

Another interesting consequence of the scenario is that the
huge amount of hot intracluster gas found today in clusters may have
been accreted only after the formation of the clusters by the
merging of many groups and the formation of most of the galaxies in
it. This is a direct consequence of the shallower potential wells of
groups. This scenario greatly alleviates the problem of extra
cooling for the formation of galaxies in clusters and may
better explain the process of metal enrichment of the intracluster
gas.

\acknowledgements This research was supported, in part, by CONACyT
grant 47282-F. The Second Palomar Observatory Sky Survey was made by
the California Institute of Technology with funds from the National
Science Foundation, the National Geographic Society, the Sloan
Foundation, the Samuel Oschin Foundation, and the Eastman Kodak
Corporation. This research has also made use of the NASA/IPAC
Extragalactic Database (NED), which is operated by the Jet
Propulsion Laboratory, California Institute of Technology, under
contract with the National Aeronautics and Space Administration and
the HyperLEDA database (leda.univ-lyon1.fr). To do the statistical
analysis and prepare the figures, we used GraphPad Prism version 5.
GraphPad Prism is a registered trademark of GraphPad Software, Inc.
San Diego California USA (www.graphpad.com). We thank M.\ Colless
and H.\ Jones for a preliminary version of the 6dF DR3 (Jones et
al.\ 2009, in prep.; see Jones et al.\ 2004 for a description of
6dF), and two anonymous referees for useful comments. We also thank
D.A.\ Andernach and R.A.~Ortega for help in identifying BCMs in a
subset of clusters.

\clearpage

\begin{figure}
\epsscale{1.0} \plotone{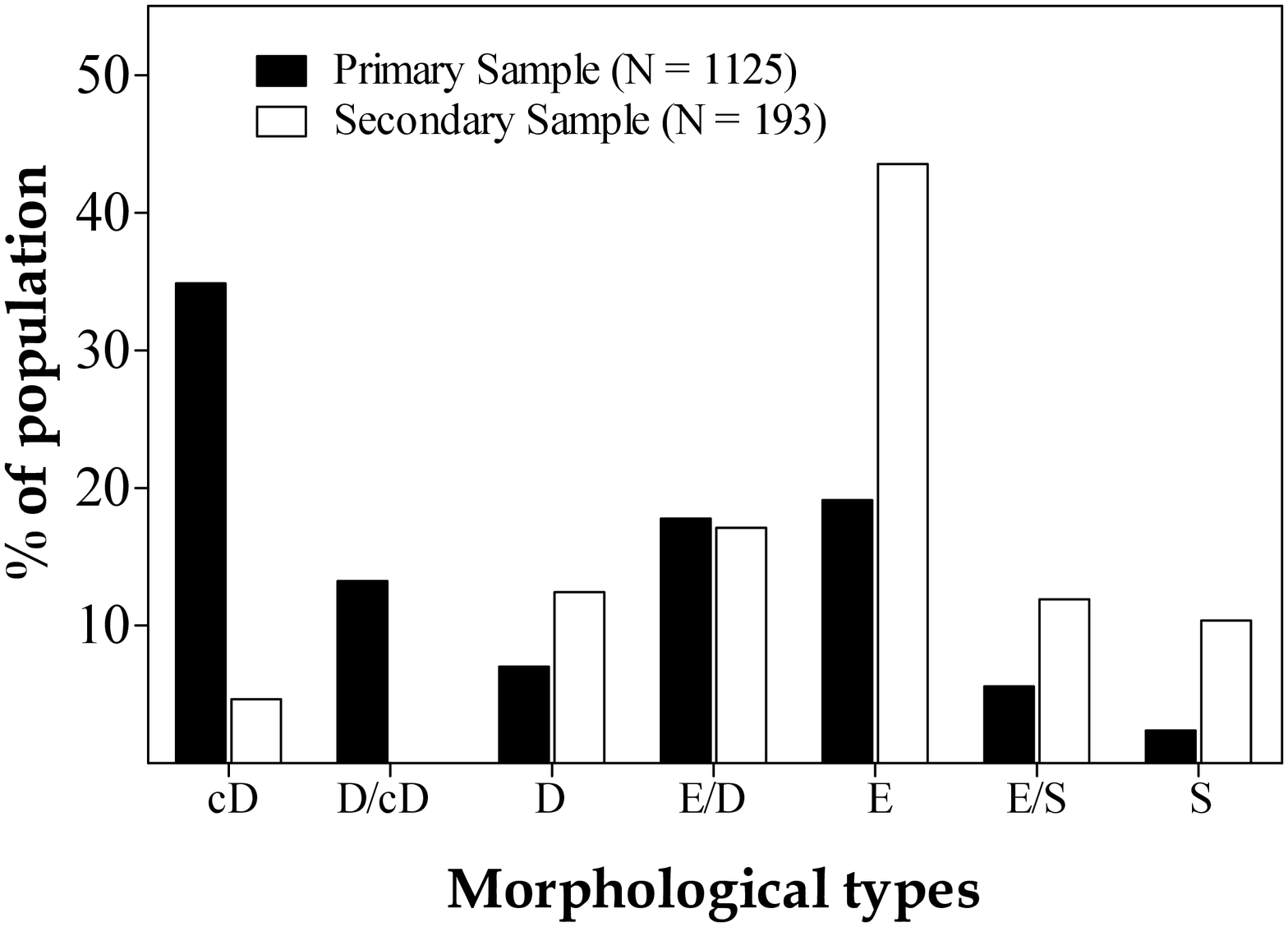} \caption{Morphology
distributions of the BCMs in our sample. The definitions of the
Primary and Secondary statistical samples are given in the text.
\label{fig1} }
\end{figure}

\clearpage

\begin{figure}
\epsscale{1.0}
\plotone{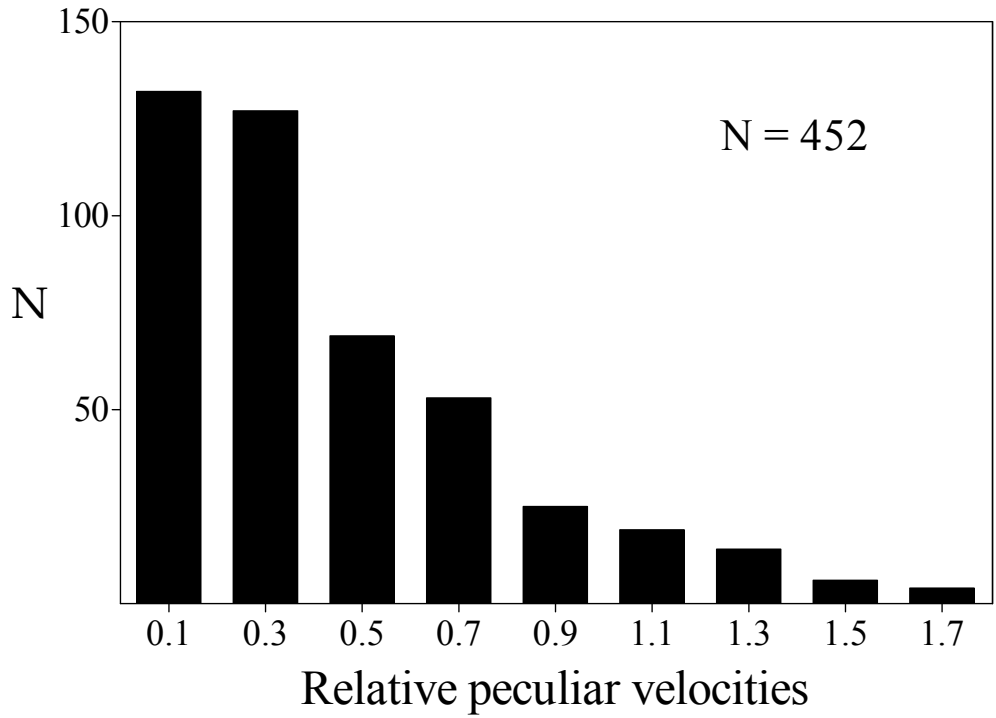}
\caption{ Distribution of the
relative peculiar velocities of the BCMs in our sample. For
practical reason, 3 BCMs with relative velocities higher than 2
are not shown.
\label{fig2} }
\end{figure}

\clearpage

\begin{figure}
\epsscale{1.0} \plotone{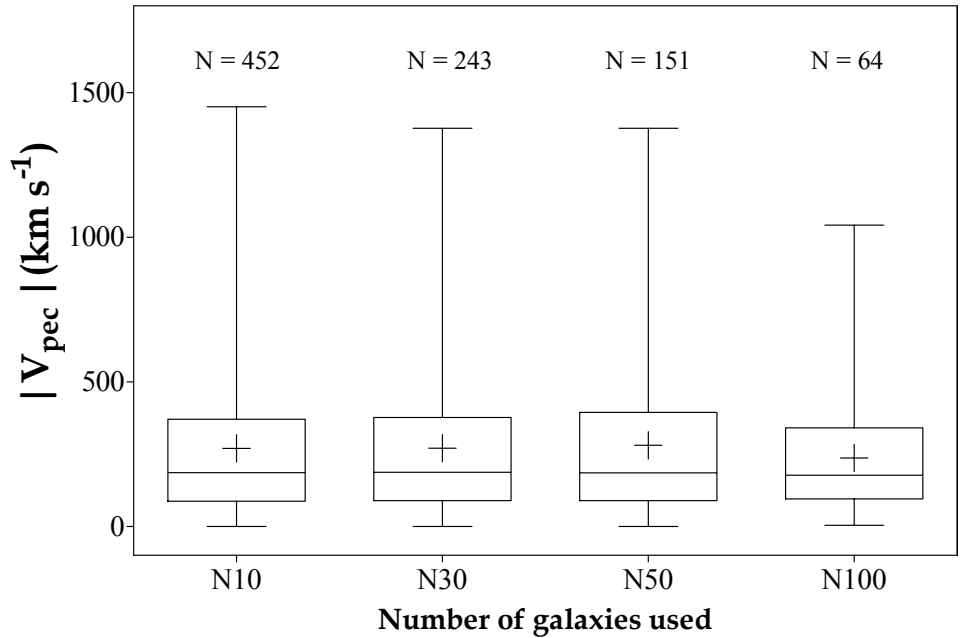} \caption {Box-whisker plots
showing the distributions of the peculiar velocities of BCMs in
samples with different minimum number of galaxies measured (N$_z$),
used to determine the cluster mean velocity and dispersion. The
definitions of the statistical samples are given in the text. The
numbers above each box indicate the sample sizes. The upper and
lower limits of the boxes are the 75\% and 25\% percentiles
respectively. The extent of the vertical bars indicate the full
range of the data (from minimum to maximum). The means are shown as
crosses at the box centers. \label{fig3}}
\end{figure}

\clearpage

\begin{figure}
\epsscale{1.0} \plotone{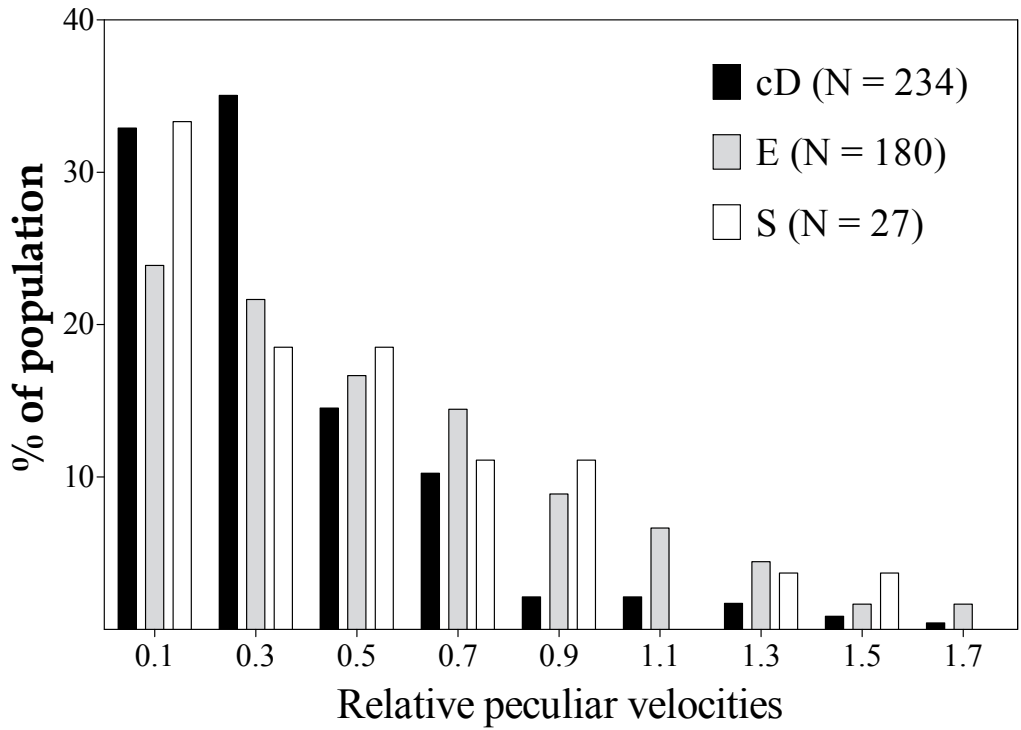} \caption{The distributions of
the relative peculiar velocities of BCMs in subsamples separated by
morphological types. 3 BCMs (2 cDs and 1 E) with relative velocities
higher than 2 are not shown. \label{fig4} }
\end{figure}

\clearpage

\begin{figure}
\epsscale{1.0} \plotone{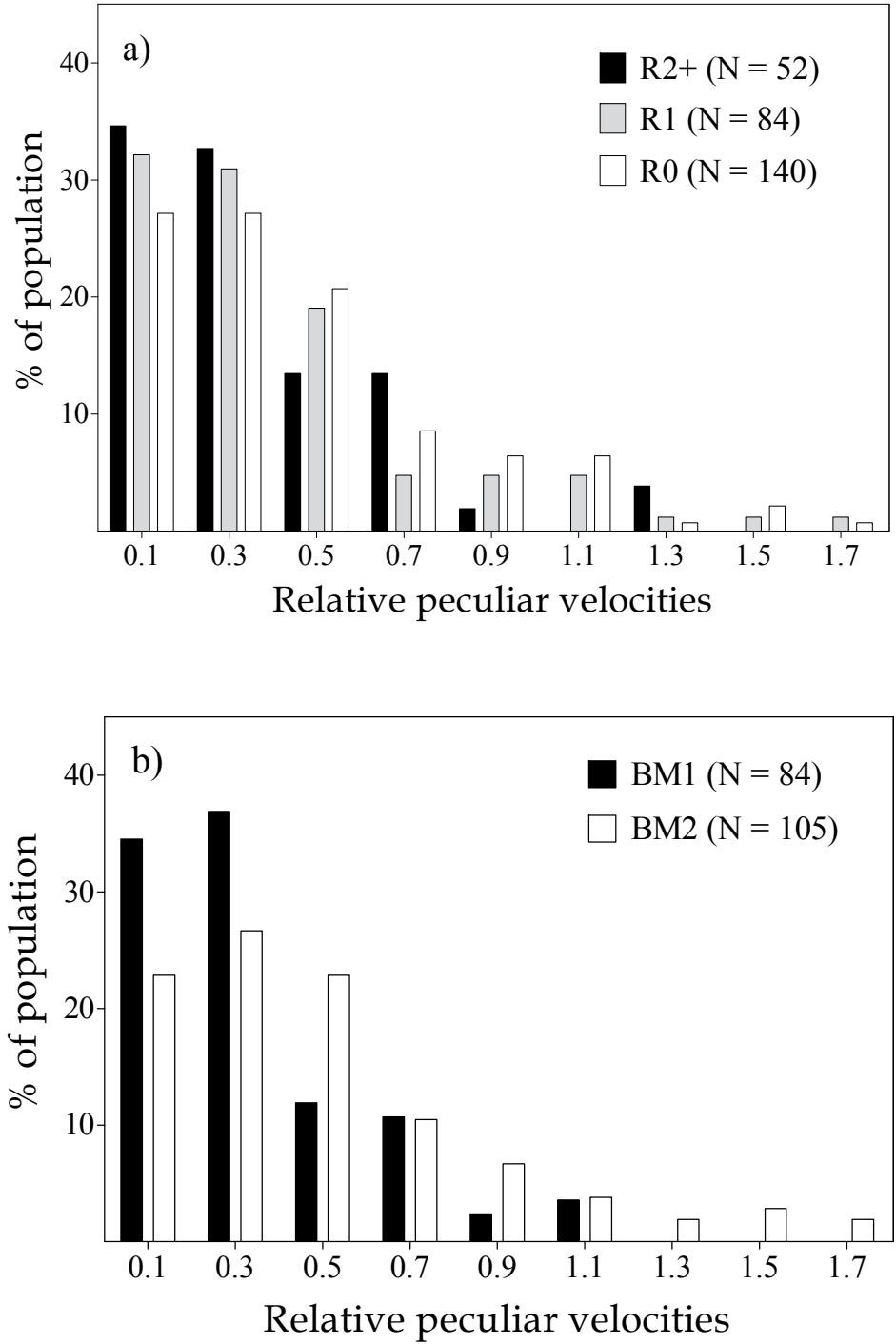} \caption{Distributions of
relative peculiar velocities of BCM in subsamples a) separated by
richness; b) separated by BM types. The definitions of the
statistical samples are given in the text. \label{fig5}}
\end{figure}

\clearpage

\begin{figure}
\epsscale{1.0} \plotone{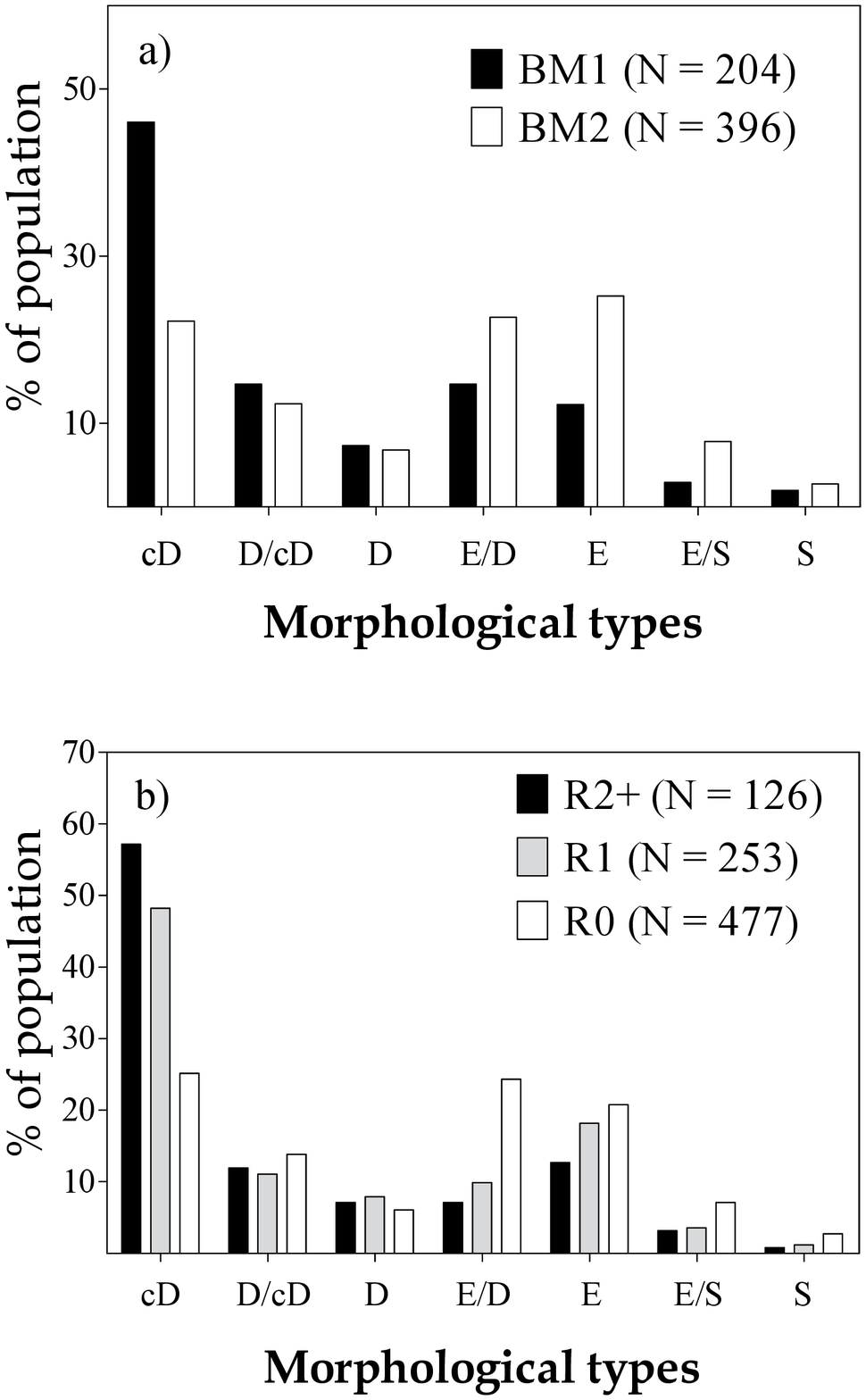} \caption{ a) Distribution of
the BCM morphologies in clusters having different BM types; b)
Distribution of the BCM morphologies in clusters having different
richness. \label{fig6}}
\end{figure}

\clearpage

\begin{figure}
\epsscale{1.0} \plotone{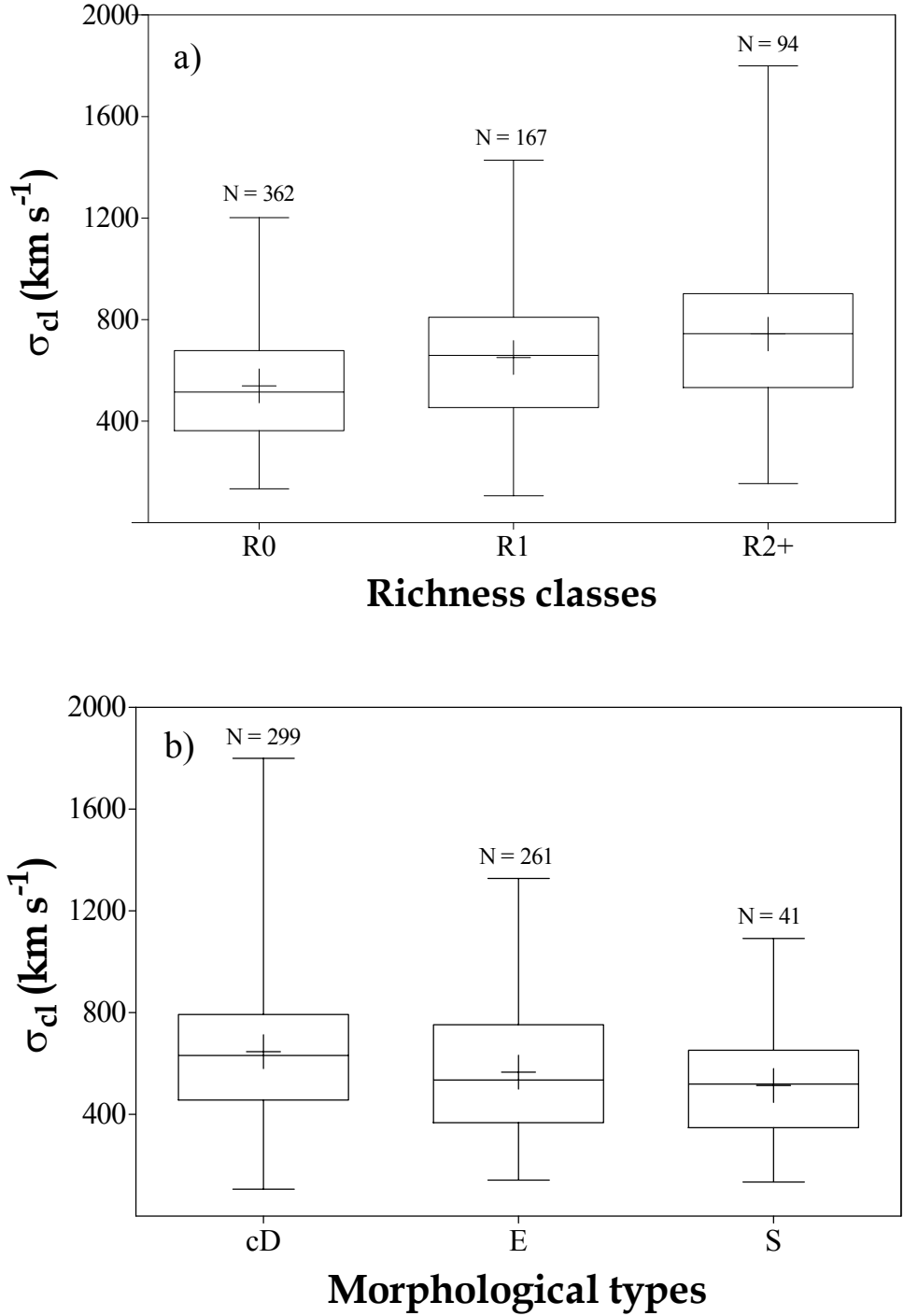} \caption {Box-whisker plots for
the velocity dispersions of galaxies in subsamples a) separated by
richness; b) separated by BCM morphology. The numbers above each box
indicate the sample sizes. The layout is the same as in figure 3.
\label{fig7}}
\end{figure}

\clearpage

\begin{figure}
\epsscale{1.0} \plotone{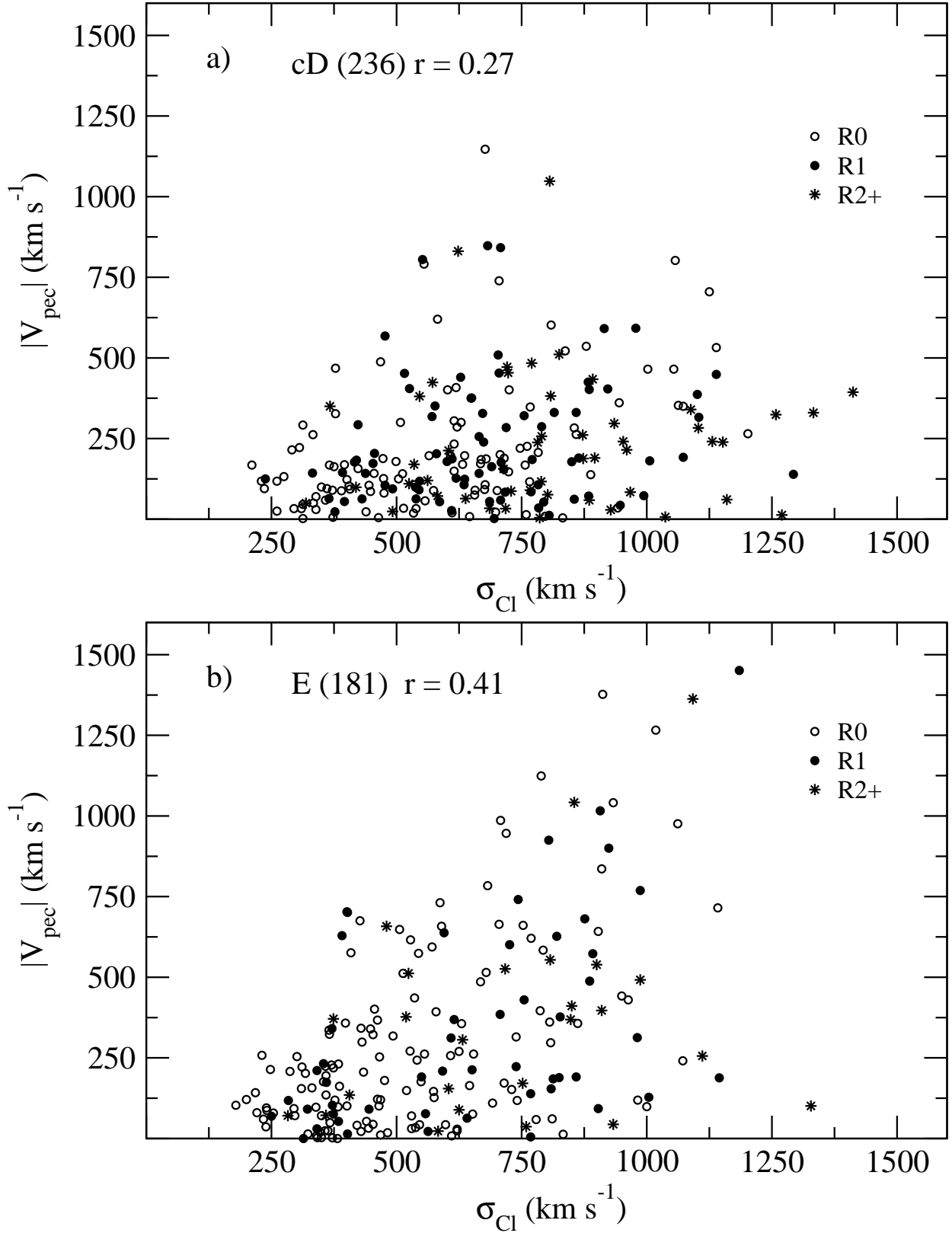} \caption {Relation between the
peculiar velocity and velocity dispersion. In a) we treat the case
of the cD subsample. In b) we show the case for the E subsample. The
different richness classes (as defined before) are indicated by
different symbols. For both subsample, we also give the number of
BCMs and the correlation coefficient, r, both with
$P<0.0001$.\label{fig9}}
\end{figure}

\clearpage

\begin{figure}
\epsscale{1.0} \plotone{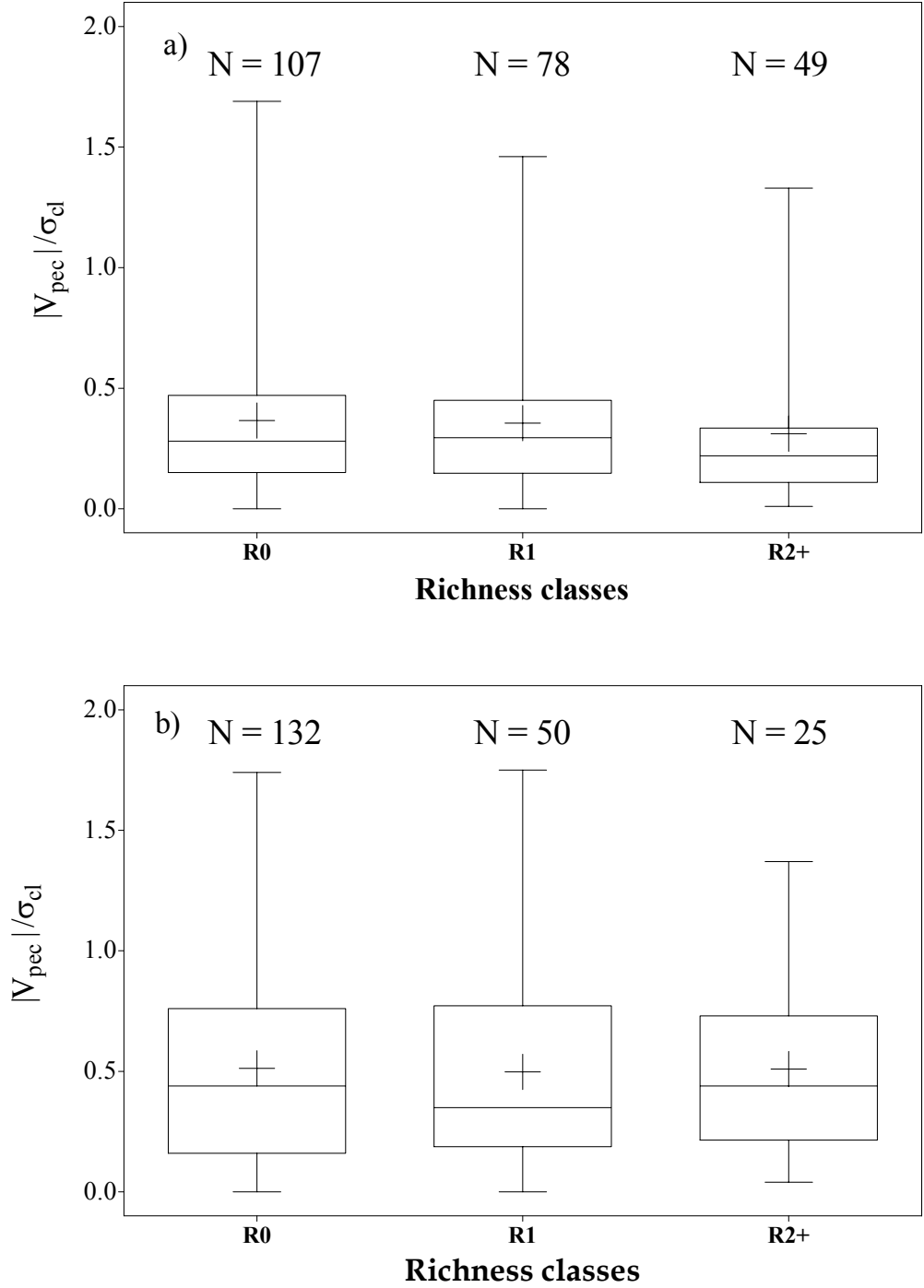} \caption{Box-whisker plots for
the relative peculiar velocity for a) the cD subsample b) the E
subsample. The layout is the same as in figure 3. \label{fig8}}
\end{figure}

\clearpage

\clearpage

\begin{deluxetable}{lccccccccc}
\tabletypesize{\scriptsize} \tablecaption{Identification of the
BCM's in their cluster}
\tablewidth{0pt}
\tablehead{
\colhead{Abell No.} & \colhead{ R. A.}    & \colhead{ Dec.}      & \colhead{R}   & \colhead{BM\tablenotemark{a}} & \colhead{RS/ACO89}  & \colhead{$v_{cl}$\tablenotemark{b}}       & \colhead{$N_z$}& \colhead{$\sigma_{cl}$\tablenotemark{c}}  & \colhead{$v_{BCM}$\tablenotemark{d}}      \\
\colhead{}          & \colhead{$(J2000)$} & \colhead{$(J2000)$}  & \colhead{}    & \colhead{}   & \colhead{}    & \colhead{$(km\ s^{-1})$} & \colhead{}     & \colhead{$(km\ s^{-1})$} & \colhead{$(km\ s^{-1})$} \\
\colhead{(1)}       & \colhead{(2)}       & \colhead{(3)}        & \colhead{(4)} & \colhead{(5)}& \colhead{(6)} & \colhead{(7)}            & \colhead{(8)}  & \colhead{(9)}            & \colhead{(10)} }
\startdata
A0002   & $00\ 08\ 16.87$ & $-19\ 39\ 41.6$ & 1 &(1:)&  B     & 36620   &    2 &       & 36720  \\
A0005   & $00\ 10\ 09.09$ & $+33\ 07\ 16.4$ & 1 &    & cD     & 41000e  &      &       &        \\
A0017   & $00\ 17\ 06.38$ & $+08\ 49\ 44.9$ & 1 & 2  & cD     & 26961n  &    2 &       & 26440  \\
A0021   & $00\ 20\ 37.19$ & $+28\ 39\ 33.6$ & 1 & 1: &  B  b  & 28401   &   15 &  855  & 28333  \\
A0022B  & $00\ 20\ 43.11$ & $-25\ 42\ 28.4$ & 3 &(2) &  B  b  & 42432   &   77 &  790  & 42725  \\
A0034A  & $00\ 27\ 33.30$ & $-08\ 53\ 11.4$ & 2 &(2) &  I  c  & 39657   &    2 &       & 39333* \\
A0034B  & $00\ 27\ 04.76$ & $-08\ 47\ 03.3$ & 2 &(2) &  I  c  & 56535   &    2 &       & 56493* \\
A0035   & $00\ 27\ 23.47$ & $-21\ 33\ 01.5$ & 3 & 5: &  I  c  & 58000e  &      &       &        \\
\enddata
\tablenotetext{a}{BM types were converted from roman numbers I, I-II, ... III,
to 1, 2, ... 5, and are given in parenthesis if this cluster has
a discarded BCM candidate in Table~3.} 
\tablenotetext{b}{ Appended with (n) if taken from literature, 
(e) if photometric estimate, (:) if poor spectrum, 
(*) factor 2 to 4 different from photometric estimate.} 
\tablenotetext{c}{ Appended with (n) if taken from literature.} 
\tablenotetext{d}{ Appended with (*) if taken from Andernach \& Tago's database.}
\end{deluxetable}

\clearpage

\begin{deluxetable}{lccccc}
\tabletypesize{\scriptsize} \tablecaption{Properties of the BCMs}
\tablewidth{0pt} \tablehead{
\colhead{Abell No.} & \colhead{$v_{pec}$}      & \colhead{$v_{pec}/\sigma_{cl}$}& \colhead{Morph.}  & \colhead{2MASX} & \colhead{Other names} \\
\colhead{}          & \colhead{$(km\ s^{-1})$} & \colhead{}                     & \colhead{}        & \colhead{}      & \colhead{} \\
\colhead{(1)}       & \colhead{(2)}            & \colhead{(3)}                  & \colhead{(4)}     & \colhead{(5)}   & \colhead{(6)} }
\startdata
A0002   &       &       & D      &  J00081685-1939423   &                                        \\
A0005   &       &       & cD     &  J00100909+3307162   &                                        \\
A0017   &       &       & cD     &  J00170632+0849445   &                                        \\
A0021   &   -62 & -0.07 & cD     &                      & MAPS-PP E 1257 0630874                 \\
A0022B  &   257 &  0.32 & cD db  &  J00204314-2542284ID &                                        \\
A0034A  &       &       & D/cD   &  J00273334-0853112   &                                        \\
A0034B  &       &       & D      &                      & SDSS J002704.78-084703.4               \\
A0035   &       &       & D int  &  J00272350-2133017   &                                        \\
\enddata
\end{deluxetable}

\clearpage

\clearpage

\begin{deluxetable}{lccccc}
\tabletypesize{\scriptsize} \tablecaption{Identification of discarded BCMs and the clusters on which they are projected.}
\tablewidth{0pt}
\tablehead{
\colhead{Abell No.} & \colhead{ R. A.}    & \colhead{ Dec.}      & \colhead{$v_{cl}$\tablenotemark{a}}       & \colhead{$v_{BCM}$\tablenotemark{b}}      & \colhead{Other names}    \\
\colhead{}          & \colhead{$(J2000)$} & \colhead{$(J2000)$}  & \colhead{$(km\ s^{-1})$}                  & \colhead{$(km\ s^{-1})$}                  & \colhead{}               \\
\colhead{(1)}       & \colhead{(2)}       & \colhead{(3)}        & \colhead{(4)}                             & \colhead{(5)}                             & \colhead{(6)} }
\startdata
A0002  & $00\ 07\ 50.56$& $-19\ 42\ 26.8$  & 36575     & 19551    & 2MASX J00075059-1942265                   \\
A0022  & $00\ 20\ 39.10$& $-25\ 35\ 29.7$  & 38548     & 19097    & 2dFGRS S138Z189                           \\
A0034  & $00\ 27\ 19.40$& $-08\ 48\ 21.0$  & 39657     & 12193    & 2MASX J00271944-0848212                   \\
A0099  & $00\ 45\ 11.38$& $-17\ 16\ 10.1$  & 28794n    &  8562    & 2MASX J00451141-1716107                   \\
A0107  & $00\ 50\ 15.33$& $-19\ 14\ 43.0$  & 17952n    & 25809~b  & 2MASX J00501533-1914431                   \\
A0180  & $01\ 22\ 31.19$& $+03\ 07\ 31.0$  & 40368n    & 35961    & 2MASX J01223119+0307312                   \\
A0327  & $02\ 12\ 43.64$& $-26\ 09\ 49.9$  & 50801     & 17628    & 2MASX J02124363-2609504                   \\
A0403  & $02\ 58\ 10.24$& $+03\ 21\ 42.4$  & 30044     &  3126    & NGC 1153                                  \\
\enddata
\tablenotetext{a}{ Only one radial velocity for a representative cluster component is given for 
clusters with known superposition; we made sure that $v_{BCM}$ is incompatible with all redshift
components known to us; appended with ``n'' if taken from literature, ``:'' if uncertain, and ``e'' 
if photometric estimate; an asterisk indicates a value that is a factor of 2 to 4 lower than 
the photometric estimate.}
\tablenotetext{b}{ Appended with an asterisk if taken from Andernach's database, and with ``b'' if
in the background to the cluster.}
\end{deluxetable}

\clearpage

\begin{deluxetable}{ll}
\tabletypesize{\scriptsize} \tablecaption{List of discarded clusters}
\tablewidth{0pt}
\tablehead{
\colhead{Abell No.} & \colhead{Reason}         \\
\colhead{(1)}       & \colhead{(2)} }
\startdata
A0319 & DSS plate defect         \\
A0812 & BCM is bg         \\
A1329 & BCM is fg         \\
A1727 & no dominant galaxy; faint         \\
A2023 & BCM is fg         \\
A2724 & no dominant galaxy; faint         \\
A2725 & too faint         \\
A2896 & BCM is fg         \\
A2942 & BCM is fg         \\
A2944 & BCM is fg         \\
A3052 & complex field; cluster not BM1         \\
A3056 & BCM is fg         \\
A3087 & BCMs are fg/bg         \\
A3307 & BCM is fg         \\
A3428 & BCM is fg; very sparse         \\
A3459 & BCM is fg         \\
A3462 & BCM is fg         \\
A3542 & BCM is fg         \\
A3599 & BCM is fg         \\
A3602 & BCM is fg; low gal. latitude         \\
A3677 & BCM in between two subcl. along line of sight \\
A3778 & no dominant galaxy         \\
A3833 & same as A3832         \\
S0294 & BCMs are fg/bg         \\
S0329 & BCM is fg         \\
S0372 & BCM is fg        \\
S0376 & no dominant galaxy         \\
S0379 & BCM is fg         \\
S0571 & BCM is fg         \\
S0596 & too faint         \\
S0631 & BCM is fg         \\
S0675 & BCM is bg         \\
S0698 & BCM is fg         \\
S0715 & BCM is bg         \\
S0790 & too faint, low galact. lat.         \\
S0812 & BCM probably fg         \\
S0852 & BCM is fg         \\
S0933 & no dominant galaxy \\
\enddata
\end{deluxetable}

\clearpage
\begin{deluxetable}{lcccc}
\tabletypesize{\scriptsize} \tablecaption{Relative peculiar velocities of the BCMs as function of their morphological type.}
\tablewidth{0pt}
\tablehead{
\colhead{Samples:}   & \colhead{All}  & \colhead{cD}    & \colhead{E}    & \colhead{S} \\
}
\startdata
Number of BCMs:                  &  452&  234 & 180  &   27 \\
percentile 25                    & 0.15& 0.15 & 0.20 & 0.14 \\
Median $|v_{pec}|/{\sigma}_{cl}$  & 0.32& 0.27 & 0.45 & 0.38 \\
percentile 75                    & 0.63& 0.46 & 0.76 & 0.66 \\
Mean  $|v_{pec}|/{\sigma}_{cl}$   & 0.44& 0.35 & 0.52 & 0.43 \\
Std. error on mean:              & 0.02& 0.02 & 0.03 & 0.07 \\
\enddata
\end{deluxetable}\clearpage

\begin{deluxetable}{lccccc}
\tabletypesize{\scriptsize}
\tablecaption{Relative peculiar velocities of the BCMs vs.\ cluster richness and BM type}
\tablewidth{0pt}
\tablehead{
\colhead{Samples:}   & \colhead{R0}  & \colhead{R1}    & \colhead{R2+}   & \colhead{BM1}      & \colhead{BM2}  \\
}
\startdata

Number of BCMs:                    &  140 &    84 &   52 &   84 &  105  \\
percentile 25                      & 0.16 &  0.18 & 0.09 & 0.09 & 0.21  \\
Median $|v_{pec}|/{\sigma}_{cl}$    & 0.33 &  0.32 & 0.25 & 0.26 & 0.40  \\
percentile 75                      & 0.58 &  0.52 & 0.48 & 0.48 & 0.64  \\
Mean $|v_{pec}|/{\sigma}_{cl}$      & 0.43 &  0.40 & 0.33 & 0.31 & 0.48  \\
Std. error on mean:                & 0.03 &  0.04 & 0.04 & 0.03 & 0.04  \\
\enddata
\end{deluxetable}

\begin{deluxetable}{lcccccccc}
\tabletypesize{\scriptsize} \tablecaption{Cluster velocity dispersion vs.\ cluster richness and BCM morphological type.}
\tablewidth{0pt} \tablehead{ \colhead{} & \colhead{} & \multicolumn{3}{c}{ Cluster } & \colhead{} & \multicolumn{3}{c}{ BCM type } \\
\cline{3-5} \cline{7-9} \\[-0.3 truecm]
\colhead{Samples:} & \colhead{} & \colhead{R0} & \colhead{R1} & \colhead{R2+} & \colhead{} & \colhead{cD}   & \colhead{E}    & \colhead{S} }
\startdata
Number of BCMs:                        & & 362  & 167  &  94 & & 299 &261 &  41 \\
percentile 25                          & & 362  & 453  & 532 & & 456 &368 & 348 \\
Median ${\sigma}_{cl}$ ($km\ s^{-1}$)  & & 515  & 659  & 744 & & 632 &536 & 520 \\
percentile 75                          & & 678  & 809  & 902 & & 793 &753 & 653 \\
Mean  ${\sigma}_{cl}$  ($km\ s^{-1}$)  & & 539  & 651  & 744 & & 647 &566 & 514 \\
Std. error on mean: ($km\ s^{-1}$)     & &  12  &  18  &  30 & &  15 & 15 &  33 \\
\enddata
\end{deluxetable}

\begin{deluxetable}{lcccccccc}
\tabletypesize{\scriptsize} \tablecaption{Relative peculiar
velocities of the BCMs vs.\
cluster richness}
\tablewidth{0pt}
\tablehead{ \colhead{} & \colhead{} & \multicolumn{3}{c}{ cD subsample} & \colhead{} & \multicolumn{3}{c}{ E subsample} \\
\cline{3-5} \cline{7-9} \\[-0.3 truecm]
\colhead{Richness classes} & \colhead{} & \colhead{R0} & \colhead{R1} &
\colhead{R2+} & \colhead{} & \colhead{R0}   & \colhead{R1}    &
\colhead{R2+} } \startdata
Number of BCMs:                        & & 107   & 78    &  49  & & 132  & 50   &  25 \\
percentile 25                          & & 0.15  & 0.15  & 0.11 & & 0.16 & 0.19 & 0.22 \\
Median $|v_{pec}|/{\sigma}_{cl}$       & & 0.28  & 0.30  & 0.22 & & 0.44 & 0.35 & 0.44 \\
percentile 75                          & & 0.47  & 0.45  & 0.34 & & 0.76 & 0.77 & 0.73 \\
Mean $|v_{pec}|/{\sigma}_{cl}$         & & 0.37  & 0.36  & 0.31 & & 0.51 & 0.50 & 0.51 \\
Std. error on mean:                    & & 0.03  & 0.03  & 0.04 & & 0.04 & 0.06 & 0.08 \\
\enddata
\end{deluxetable}

\end{document}